\documentclass[fleqn,usenatbib]{mnras}

\usepackage{newtxtext,newtxmath}

\usepackage[T1]{fontenc}
\usepackage{ae,aecompl}

\usepackage{graphicx}	
\usepackage{amsmath}	
\usepackage{amssymb}	

\newcommand{\kms}{$\mathrm{km\,s^{-1}}$}
\newcommand\Msol{M$_{\odot}$}
\newcommand\Lsol{L$_{\odot}$}

\newcommand{\hi}{H\,{\sc i}}

\title[Group \hi \ mass function]{The \hi \ mass function of group galaxies in the ALFALFA survey}

\author[Jones et al.]{Michael G. Jones$^{1}$\thanks{E-mail: mjones@iaa.es},
Kelley M. Hess$^{2,3}$\thanks{E-mail: hess@astro.rug.nl},
Elizabeth A.~K. Adams$^{3,2}$ \newauthor
and Lourdes Verdes-Montenegro$^{1}$
\\
$^{1}$Instituto de Astrof\'{i}sica de Andaluc\'{i}a (CSIC), Glorieta de la Astronom\'{i}a s/n, 18008 Granada, Spain
\\
$^{2}$Kapteyn Astronomical Institute, University of Groningen, Landleven 12, Groningen NL-9747AD, The Netherlands\\
$^{3}$ASTRON, the Netherlands Institute for Radio Astronomy, Postbus 2, Dwingeloo NL-7900AA, The Netherlands
}

\date{Accepted XXX. Received YYY; in original form ZZZ}

\pubyear{2020}

\begin{document}
\label{firstpage}
\pagerange{\pageref{firstpage}--\pageref{lastpage}}
\maketitle

\begin{abstract}
We estimate the \hi \ mass function (HIMF) of galaxies in groups based on thousands of ALFALFA (Arecibo Legacy Fast ALFA survey) \hi \ detections within the galaxy groups of four widely used SDSS (Sloan Digital Sky Survey) groups catalogues. Although differences between the catalogues mean that there is no one definitive group galaxy HIMF, in general we find that the low-mass slope is flat, in agreement with studies based on small samples of individual groups, and that the `knee' mass is slightly higher than that of the global HIMF of the full ALFALFA sample. We find that the observed fraction of ALFALFA galaxies in groups is approximately 22 per cent. These group galaxies were removed from the full ALFALFA source catalogue to calculate the field HIMF using the remaining galaxies. Comparison between the field and group HIMFs reveals that group galaxies make only a small contribution to the global HIMF as most ALFALFA galaxies are in the field, but beyond the HIMF `knee' group galaxies dominate. Finally we attempt to separate the group galaxy HIMF into bins of group halo mass, but find that too few low-mass galaxies are detected in the most massive groups to tightly constrain the slope, owing to the rarity of such groups in the nearby Universe where low-mass galaxies are detectable with existing \hi \ surveys.
 \end{abstract}

\begin{keywords}
Galaxies, mass function, radio surveys
\end{keywords}



\section{Introduction}

Over the past few decades the HIMF (\hi \ mass function; the number density of galaxies in the Universe as a function of their \hi \ mass) has gone from being highly uncertain to well known, at least in the nearby Universe, thanks mostly to the two largest blind \hi \ surveys to date, HIPASS \citep[\hi \ Parkes All Sky Survey;][]{Barnes+2001,Meyer+2004} and ALFALFA \citep[Arecibo Legacy Fast ALFA survey;][]{Giovanelli+2005,Haynes+2011,Haynes+2018}. Hence, the most recent studies of the HIMF have focused on its potential variation with environment \citep{Moorman+2014,Jones+2016b,Jones+2018,Said+2019} and have begun to push the measurement domain beyond $z \approx 0$ \citep{Hoppmann+2015}.  

In addition to blind measurements of the HIMF over wide fields \citep{Zwaan+1997,Rosenberg+2002,Zwaan+2003,Zwaan+2005,Martin+2010,Jones+2018,Said+2019}, there have been a number of measurements of the HIMF in specific galaxy groups \citep{Verheijen+2001,Kovac+2005,Freeland+2009,Kilborn+2009,Stierwalt+2009,Davies+2011,Pisano+2011,Westmeier+2017}. While the former have shown that the HIMF in the local Universe follows a Schechter function \citep{Schechter1976} shape (a declining power law with increasing \hi \ mass that is truncated by an exponential decay at the `knee' mass) with a low-mass slope parameter, $\alpha$, of approximately -1.3 and a `knee' mass just below $10^{10}$ \Msol, the latter studies have almost universally found the low-mass slope in galaxy groups to be flat ($\alpha \approx -1$) down to about $\log{M_{\mathrm{H\,\textsc{i}}}/\mathrm{M_\odot}} \sim 7$. However, these samples have typically consisted of 1-20 groups containing a total of 30-300 \hi \ detections, compared to the many thousands of detections in blind, wide-field surveys.

There is now a considerable body of evidence, both from simulations \citep{Bahe+2015,Marasco+2016,Jung+2018} and observations \citep{Hess+2013,Denes+2016,Jaffe+2016,Odekon+2016,Brown+2017}, indicating that galaxies are ``pre-processed'' in groups, depleting their HI content and suppressing their star formation rate, before they fall into clusters. Hence the difference between the HIMF found in groups and in wide-field surveys is not entirely unexpected as the \hi-rich galaxies typically detected by blind 21 cm surveys are mostly thought to be field galaxies in their own distinct halos \citep[e.g.][]{Guo+2017}, whereas most galaxies in groups are, by definition, satellites. However, it is somewhat intriguing that no conclusive evidence for a flattening slope going from low to higher density environments has been found \citep[e.g.][]{Moorman+2014,Jones+2016b}.\footnote{We note here that \citet{Said+2019} did recently find such as trend in the HIZOA (\hi \ Zone Of Avoidance) survey \citep{Staveley-Smith+2016}, but they used \hi \ neighbours to define environment and it is unclear how well this correlates with the local galaxy or total mass density. In addition, \citet{Zwaan+2005} used a similar metric but found the opposite trend. More work is needed to fully understand these results.}
Furthermore, there are a few groups that do not fit the apparent trend of the other studies and have been found to have low-mass slopes even steeper than the global measurements in HIPASS and ALFALFA \citep{Stierwalt+2009,Davies+2011}.

There appear to be two possible resolutions to these potentially conflicting results: 1) that the environment metrics used to study environmental dependence of the HIMF in wide-field surveys, did not adequately separate group and cluster environments from the field population, thereby preventing any shift in the low-mass slope from being detected, or 2) that there is a methodological inconsistency in how the wide-field survey HIMFs and those of individual groups were calculated, and the difference in $\alpha$ is not actually as large as has been reported. One potential source of inconsistency between group and field HIMFs could be that ALFALFA and HIPASS are both surveys conducted with single dish telescopes, whereas most group HIMFs have been measured using interferometric observations of individual groups. Though unlikely, this raises the possibility of differences in the completeness corrections used, for example because interferometers suffer from surface brightness sensitivity limitations, but for both Parkes and Arecibo the vast majority of galaxies are point-like when observed at 21 cm, resulting in a simpler sensitivity limit. On the other hand, there are also reasons to suspect the first potential resolution. While the void-wall and nearest neighbour environment metrics used by \citet{Moorman+2014} and \citet{Jones+2016b} were shown to separate regions of different galaxy environments on relatively large scales, it is not necessarily the case that groups would be concentrated in one region of the metrics' parameter space.

In this paper we aim to resolve this tension by using the ALFALFA source catalogue to calculate the average group galaxy HIMF based on the many groups contained in the ALFALFA survey volume. This approach sidesteps both of the issues discussed above, as the dataset was observed with a single dish and by matching ALFALFA detections to optical group catalogues we can also avoid complications with more general environment metrics. In addition, this approach results in a group galaxy sample of thousands, rather than tens or hundreds, meaning that the resulting group galaxy HIMF is one of the most robust measurement to date.

The following section briefly outlines the ALFALFA survey and section \ref{sec:group_cats} presents the four different group catalogues which we use. Section \ref{sec:HIMF_calc} describes our approach to calculating the HIMF for this dataset, our results are presented and discussed in section \ref{sec:results}, and finally we conclude in section \ref{sec:conclude}. Throughout this paper we assume $H_0 = 70 \; \mathrm{km\,s^{-1}\,Mpc^{-1}}$ and that the absolute magnitude of the Sun is 4.67 in the $r$-band. Distances are approximated as $cz_\mathrm{cmb}/H_0$.

\section{The ALFALFA survey}

The ALFALFA survey \citep{Giovanelli+2005} is a blind 21 cm radio survey covering approximately 6900 deg$^2$ of sky out to a maximum redshift of 0.06. The survey was conducted over about 4500 hours of observing time with the 305 m Arecibo telescope in Puerto Rico. It followed a double-pass drift scan observing strategy using the 7 beam ALFA (Arecibo L-band Feed Array) instrument, covering the survey area in equally spaced declination strips. The final \hi \ source catalogue \citep{Haynes+2018} contains over 30000 extragalactic \hi \ sources, 25434 of which are classified as ``code 1", meaning they are high signal-to-noise detections with extremely high reliabilities and a well-defined completeness limit \citep{Haynes+2011,Haynes+2018}. The survey area is split into two continuous regions, one in the Northern Spring sky (approximately 7.5 hr $<$ RA $<$ 16.5 hr) and one in the Fall sky (approximately 22 hr $<$ RA $<$ 3 hr), both range from 0$^\circ$ to +36$^\circ$ in declination. As the groups catalogues which we will make use of are based on SDSS spectroscopic galaxy catalogues there is only appreciable overlap with ALFALFA in the Spring portion of the survey. It is therefore the code 1 sources in the Northern Spring sky that we will use throughout this paper to make all our estimates of the group galaxy HIMF.

For any measurement of the HIMF a key quantity for each galaxy is its \hi \ mass. To estimate the \hi \ mass we use the standard expression
\begin{equation}
\frac{M_{\mathrm{H\,\textsc{i}}}}{\mathrm{M}_{\odot}} = 2.356 \times 10^{5} D_{\mathrm{Mpc}}^{2} S_{21},
\end{equation}
where $D_{\mathrm{Mpc}}$ is the distance to the galaxy in Mpc and $S_{21}$ is its integrated flux in Jy \kms. In this case we do not adopt the ALFALFA distance estimates for the galaxies assigned to groups (see below), instead we take the redshift relative to the CMB rest frame ($z_\mathrm{cmb}$) reported in each group catalogue and calculate the Hubble-Lema\^{i}tre flow\footnote{Previously referred to as the Hubble flow.} ($H_0 = 70 \; \mathrm{km\,s^{-1}\,Mpc^{-1}}$) distance to each group. The same distance is assumed for all members of a group.

\section{Group catalogues}
\label{sec:group_cats}

Several different techniques have been developed in the attempt to identify gravitationally bound collections of galaxies in large spectroscopic redshift surveys.  Among the most common in the literature are the friends-of-friends (FoF) group finding algorithm \citep[e.g.][]{Huchra+Geller1982,Eke+2004,Crook+2007,Berlind+2006,Tempel+2014} and iterative halo-based group finders \citep[e.g.][]{Yang+2007,Lu+2016,Lim+2017}. Friends-of-friends uses linking lengths: one in sky projection and one in redshift to associate nearby galaxies with one another.  These algorithms have been tested to match the halo multiplicity and halo occupation properties of galaxies in mock galaxy catalogues from N-body simulations.  FoF algorithms are elegant in their simplicity, however the choice of linking lengths is not unique, and depend on the scientific motivation \citep{Duarte+Mamon2014}.  Halo based finders are an attempt to develop a more physically motivated algorithm which use the stellar mass of a galaxy (or collection of galaxies) as a proxy for the dark matter halo mass in order to associate galaxies in common dark matter halos. 
Such algorithms start with seed groups (for example, from a FoF algorithm with a short linking length) or individual galaxies, the luminosity of the galaxies in these seeds are then used to estimate the group halo mass, from which relations between a halo mass and its size and velocity dispersion (from theory or simulations) can be used to assign probabilities that other galaxies in the vicinity are also members of the same group. Galaxies above a certain threshold probability are incorporated into the group and the steps are iterated to convergence. In this way the assignment of galaxies to groups is more grounded in a physical model, instead of reliant on, somewhat arbitrary, choices of linking lengths. For further details of this method we refer the reader to the articles cited above.

Regardless of the method used to find groups and their members, the resulting group catalogues are always either volume-limited and therefore omit low luminosity (mass) objects at all redshifts because they are not found in the entire volume, or the group catalogues are flux-limited and therefore the group statistics change over the redshift range they cover because the lowest luminosity objects are only visible nearby.  In this paper we examine the HIMF from four different popular group catalogues constructed either using FoF \citep{Berlind+2006,Berlind+2009,Tempel+2014} or the iterative halo finder \citep{Yang+2007,Lim+2017}.  Three of these group catalogues are volume-limited \citep{Yang+2007,Berlind+2009,Tempel+2014}, while the last is flux-limited \citep{Lim+2017}.

Three of the four catalogues, as published, also include halos which consist of only one or two galaxies.  Individual, well-defined groups typically need at least 10 members to reliably estimate the size and velocity dispersion, but the lowest mass groups only have a handful of members.  Therefore, to probe the low mass, loose group environment we only consider groups with at least three members.  
Of course, it should be noted that due to differences in the methodology used to construct the groups in the four catalogues, a triplet in one catalogue is not necessarily an equivalent type of object to a triplet in another catalogue.

\subsection{Berlind et al. groups}
\defcitealias{Berlind+2009}{B09}
\defcitealias{Berlind+2006}{B06}

\citet{Berlind+2009}, hereafter \citetalias{Berlind+2009}, is an application of the FoF algorithm of \citet{Berlind+2006}, hereafter \citetalias{Berlind+2006}, used on SDSS DR4 to the SDSS DR7, which provided spectroscopic coverage complementary with the full ALFALFA coverage.  Three volume-limited catalogues are available online.\footnote{The group catalogues were obtained from the website of A.~Berlind (\url{http://lss.phy.vanderbilt.edu/groups/dr7/}), hosted by Vanderbilt University, on 5th Feb. 2018, although the content remains unchanged to the date of acceptance of this work.}  We choose the group catalogue whose absolute magnitude limit in $r$-band is the faintest at -18.0.  By including the faintest galaxies, we probe the lowest mass group regime available.  The absolute magnitude limit also effectively sets an upper redshift at $z=0.042$ which is well matched to the ALFALFA selection function.  The authors set an additional lower redshift cut of $z=0.02$ below which they do not trust the SDSS DR7 photometry.
As the inner and outer limits of this catalogue were defined using the observed redshifts of the galaxies in the heliocentric frame, which do not directly correspond to (Hubble-Lema\^{i}tre flow) distances, we removed small regions on the inner and outer boundaries by redefining the redshift limits of the catalogue in the CMB reference frame \citep{Lineweaver+1996} as 6750 \kms \ and 12500 \kms, respectively. These reshift criteria cause the velocity dispersion of the Coma cluster to be truncated and we therefore eliminated it from the catalogue entirely.

\subsection{Tempel et al. groups}
\defcitealias{Tempel+2014}{T14}

\citet{Tempel+2014}, hereafter \citetalias{Tempel+2014}, provides both flux and volume-limited catalogues.  In this case, we chose the volume-limited catalogue with the same $r$-band absolute magnitude cutoff of -18.0 as \citetalias{Berlind+2009}.  The data for this group catalogue come from SDSS DR10, which has the same coverage as DR7, but the galaxy properties on which the group finder were run are based on data from updated SDSS photometric and spectroscopic pipelines.  The first major differences between the \citetalias{Berlind+2009} and \citetalias{Tempel+2014} catalogues are that \citetalias{Tempel+2014} uses different linking lengths.
The linking length is barely smaller than \citetalias{Berlind+2006} in sky projection, but larger by almost a factor of 2 in the radial direction.  Nonetheless, \citetalias{Tempel+2014} is in line with the recommendations of \citet{Duarte+Mamon2014}, and thus \citetalias{Berlind+2006} is more conservative and may miss galaxies that are falling in along the line of sight.

The second major difference is that \citetalias{Tempel+2014} uses a higher redshift cutoff of $z<0.045$, and applies no low redshift cutoff.  As a result, the \citetalias{Tempel+2014} catalogue is incomplete for bright, very nearby objects where SDSS becomes saturated \citep{Tempel+2012}, or where the SDSS pipeline has not targeted objects due to shredding, etc.  The authors attempt to overcome this and SDSS fibre collisions by adding additional redshifts from 2dFGRS (2 degree Field Galaxy Redshift Survey), 2MRS (2MASS, 2 Micron All Sky Survey, Redshift Survey), and RC3 (3rd Reference Catalogue of bright galaxies).  Nonetheless, in fig. 2 of \citetalias{Tempel+2014} there appears to be a discontinuity in the number density of galaxies at redshift of around 4000 \kms \ ($\sim$60 Mpc) which might be indicative of incompleteness at low redshift.

In the case of the \citetalias{Tempel+2014} groups, the minimum redshift corresponds to a CMB frame recession velocity of just over 1000 \kms, so we make a conservative choice and restrict the catalogue to $1500 < cz_{\mathrm{cmb}}/\mathrm{km\,s^{-1}} < 13500$.

\subsection{Yang et al. groups}
\defcitealias{Yang+2012}{Y12}
\defcitealias{Yang+2007}{Y07}

The \citet{Yang+2012}, hereafter \citetalias{Yang+2012}, catalogue is an update of the iterative halo based group finder of \citet{Yang+2007}, hereafter \citetalias{Yang+2007}, from DR4 to the SDSS DR7 data, augmented with redshifts from 2dFGRS, 2MRS, and R3C as with \citetalias{Tempel+2014}. This is the so-called ``modelB'' catalog.  The reference to the new catalogue barely appears as a footnote in Appendix B of \citetalias{Yang+2012}, but the catalogue is available online\footnote{http://gax.sjtu.edu.cn/data/Group.html.  Note that this is the new host website to replace the previously published location (S.~Lim, private communication).}.

This catalogue differs from FoF significantly in the philosophy of its construction as well as the types of galaxies and groups it includes.  The seed galaxies for the group halos are effectively volume-limited but are required to have an $r$-band absolute magnitude limit of -19.5.  In addition, only galaxies above this limit are used to estimate the group halo mass and radius, which is in turn used to assign new members.  After the initial seeds are identified, the spectroscopic catalogue used to populate the groups is flux-limited, and any galaxy within the estimated size of the halo is included in the group, including galaxies fainter than -19.5 magnitudes in $r$-band.  Despite this, the \citetalias{Yang+2012} catalogue does not include low mass groups because galaxies fainter than -19.5 magnitudes cannot be grouped together: if they do not reside next to a bright galaxy, then they are treated as centrals in their own halos.  As a result, \citetalias{Yang+2012} groups with low membership will on average be higher mass than groups with the same membership in the \citetalias{Berlind+2009} or \citetalias{Tempel+2014} catalogues.

As with the \citetalias{Berlind+2006} groups the \citetalias{Yang+2007} groups have inner and outer boundaries based on heliocentric redshifts. In this case we restrict the catalogue to $3600 < cz_{\mathrm{cmb}}/\mathrm{km\,s^{-1}} < 15000$.

\subsection{Lim et al. groups}
\defcitealias{Lim+2017}{L17}

\citet{Lim+2017}, hereafter \citetalias{Lim+2017}, is a improvement on the methodology developed by \citetalias{Yang+2007} to extend it to poor and low mass groups.  The paper provides group catalogues for four spectroscopic surveys: 2MRS, 6dFGS (6 degree Field Galaxy Survey), SDSS and 2dFGRS. For its depth and completeness, we use their group catalogue for SDSS DR13, so-called ``SDSS+M''.  In addition, to improved photometry and spectroscopy over DR7, some of the fibre collision galaxies now have spectroscopy in DR13, and additional objects have spectra from the BOSS SDSS survey.  As with the \citetalias{Yang+2012} and \citetalias{Tempel+2014} catalogues, \citetalias{Lim+2017} augments SDSS objects without redshifts with spectroscopy from other surveys including 2dFGRS, 6dFGS, KIAS VAGC (Korea Institute for Advanced Study Value Added Galaxy Catalog), LAMOST (Large sky Area Multi-Object Fiber Spectroscopic Telescope). These augments result in total redshift completeness for galaxies with apparent magnitudes brighter than 17.7 in the $r$-band. The catalogue is available online for download (see previous footnote).

In this case there is no magnitude limit applied to the galaxy catalogue and a preliminary halo mass is assigned to every galaxy.  For each halo, the size and line-of-sight velocity dispersion is calculated based on the mass of the halo (from the stellar luminosity), and the phase space distribution of galaxies in dark matter halos is used to associate galaxies into groups.  Groups are ranked and assigned masses by halo abundance matching, and the process is iterated again.

At high group masses \citetalias{Lim+2017} and \citetalias{Yang+2012} catalogues have no significant difference, but critically, the \citetalias{Lim+2017} catalogue extends to lower group masses where the abundance of groups is greater.  For studies of the impact of environment on something as tenuous as gas disks, a catalogue which extends to lower group masses is critical to understand where environmental effects become important.

As the \citetalias{Lim+2017} catalogue is not volume-limited we decided to create a sub-catalogue that would be complete (in a volume-limited sense) for all the group centrals. We set redshift boundaries as $1000 < cz_{\mathrm{cmb}}/\mathrm{km\,s^{-1}} < 13500$ and used the SDSS spectroscopic survey completeness threshold magnitude (17.7 in $r$-band) to remove any groups with centrals less luminous than the corresponding absolute magnitude at the outer distance boundary, -18.7.

\subsection{Assignment of ALFALFA sources to groups}

We assign \hi\ detected galaxies from ALFALFA to galaxy groups in a two step process:  first by directly matching them using SDSS object IDs to optical galaxies which are known to be group members; and second through a proximity matching to groups if the \hi\ detections fall within the group volume, defined by a velocity range and projected radius.  The direct matching ensures that known optical members with an \hi\ counterpart are included in the group catalogue.  The proximity matching step allows us to include gas rich galaxies which may not have a stellar counterpart in the group catalogue because they were too optically faint (for example in the flux-limited group catalogues).

\subsubsection{Direct matching}

When ALFALFA \hi\ detections were extracted during the data reduction process the most likely optical counterparts were identified manually \citep[see][section 4.1]{Haynes+2011}. These manual identifications are then automatically matched to SDSS photometric and spectroscopic objects where they are available\footnote{The SDSS cross-match in \citet{Haynes+2011} was based on DR7, while the current version of the ALFALFA crossmatch uses the objIDs and specObjIDs corresponding to DR8 onward (DR8+).}, providing a catalogue of counterparts (Durbala et al. in prep.).

The \citetalias{Berlind+2009} and \citetalias{Yang+2012} catalogues were derived from DR7, so as a first step we retrieved updated object IDs from DR8+ through the SDSS CasJobs.  In less than 1\% of cases, we did not find a DR8+ counterpart for a DR7 group member.  Usually this was because one galaxy had been shredded into multiple sources in DR7, and this issue had been fixed in the later SDSS photometric pipeline.  Given the small number of sources to be effected, we do not expect this issue to have a strong impact on the group statistics. The \citetalias{Tempel+2014} and \citetalias{Lim+2017} catalogues were derived from later SDSS data releases and their object IDs could be used as provided in the group catalogues. 

For each optical group catalogue, we matched the specObjIDs of ALFALFA detections to specObjIDs of the optical members using \texttt{join} in astropy.table.  This resulted in four tables of \hi\ ``direct matches'', one for each group catalogue.  

\subsubsection{Proximity matching}
\label{sec:prox_match}

For the proximity matching we calculated the volume of every group based on their published physical properties, and assigned an \hi\ galaxy to a given group if it fell within that volume and was not previously assigned to a group by direct matching.  To be clear: it is not necessary in this case for the \hi\ detection to have an optical spectroscopic counterpart, it simply needs to fall within the group environment to be included based on the \hi\ position and redshift.  However, we note that all of the \hi\ detections in our subset of ALFALFA at least have a photometric counterpart in SDSS. 

\begin{figure*}
\includegraphics[width=\columnwidth]{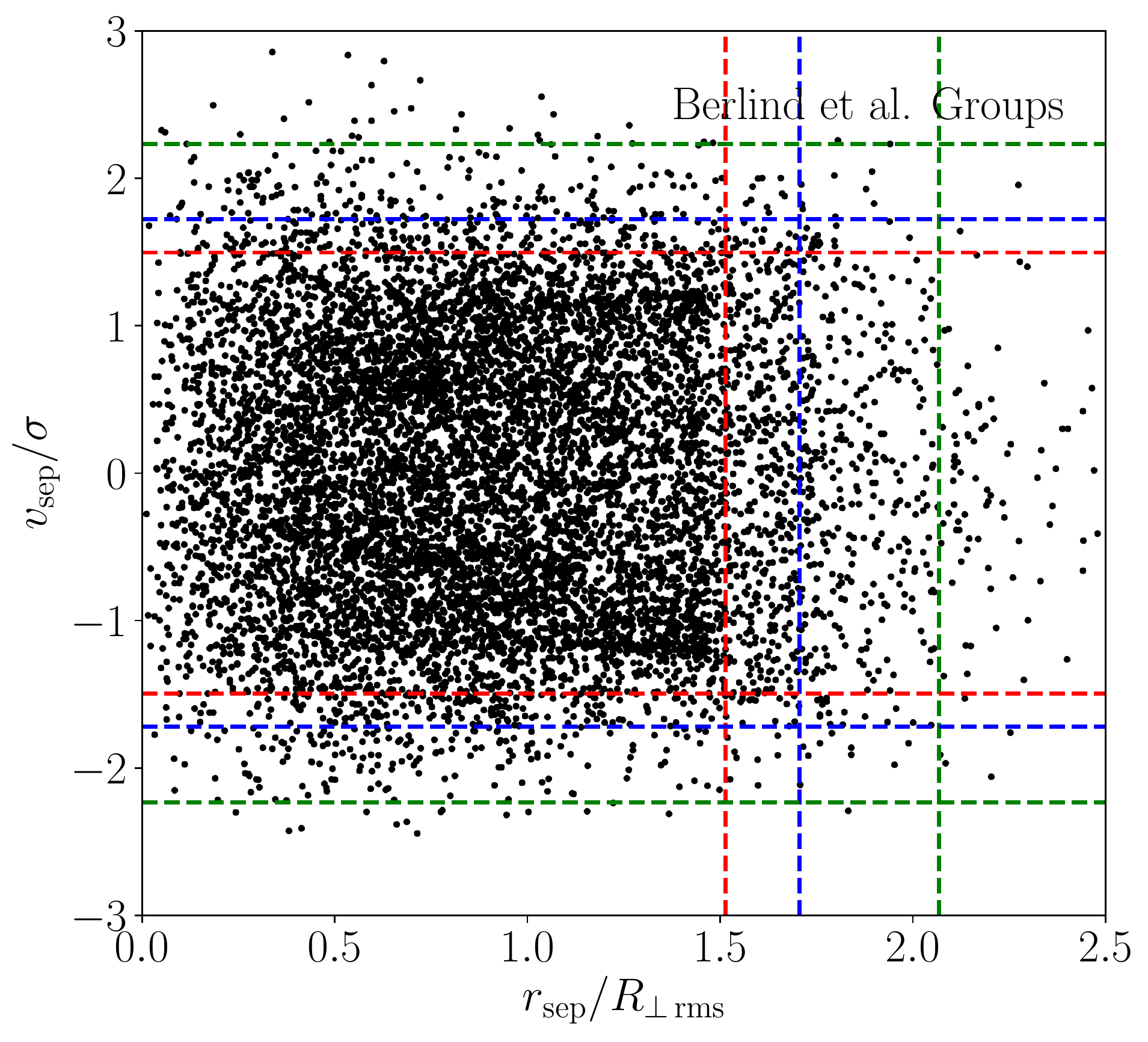}
\includegraphics[width=\columnwidth]{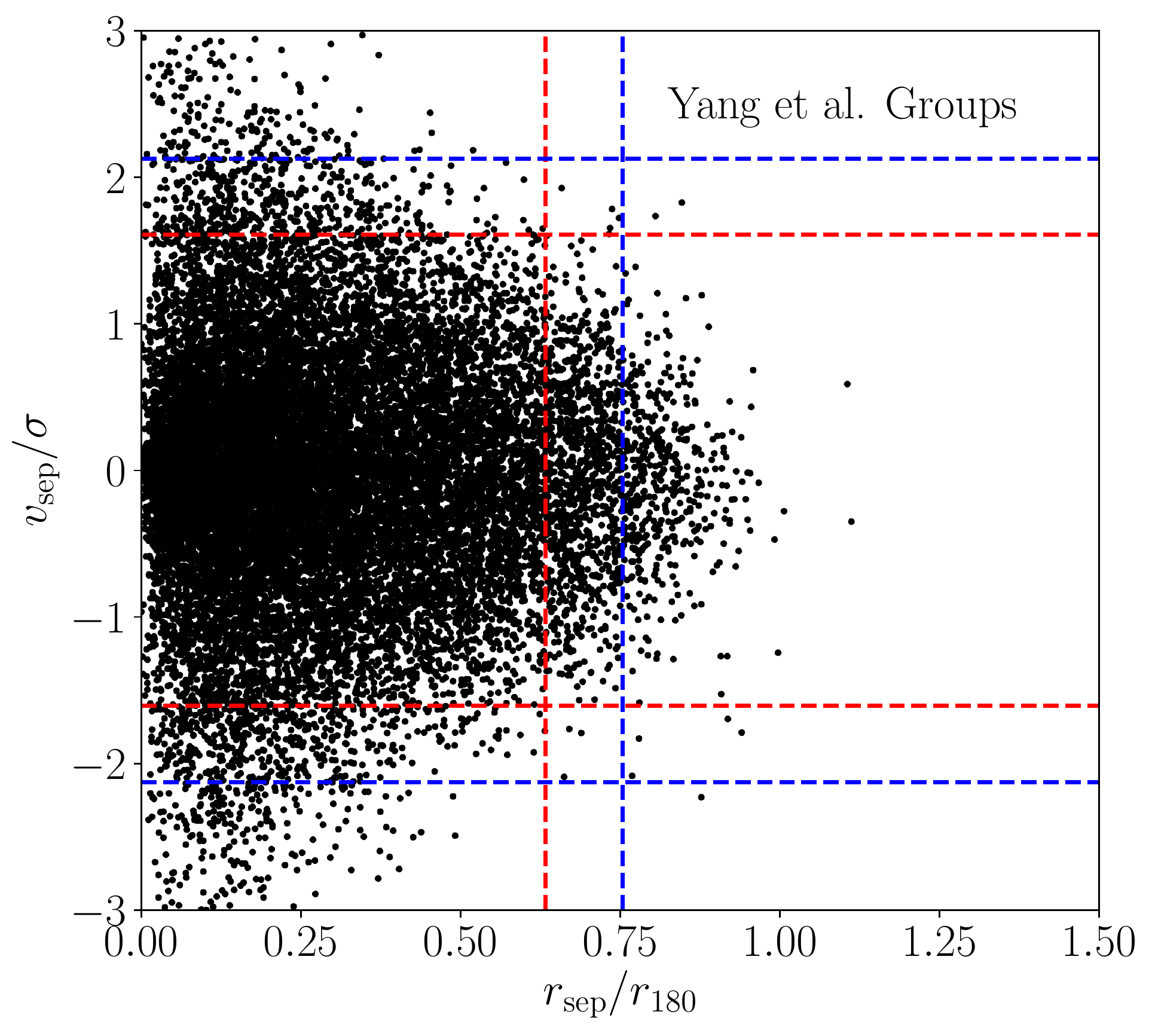}
\includegraphics[width=\columnwidth]{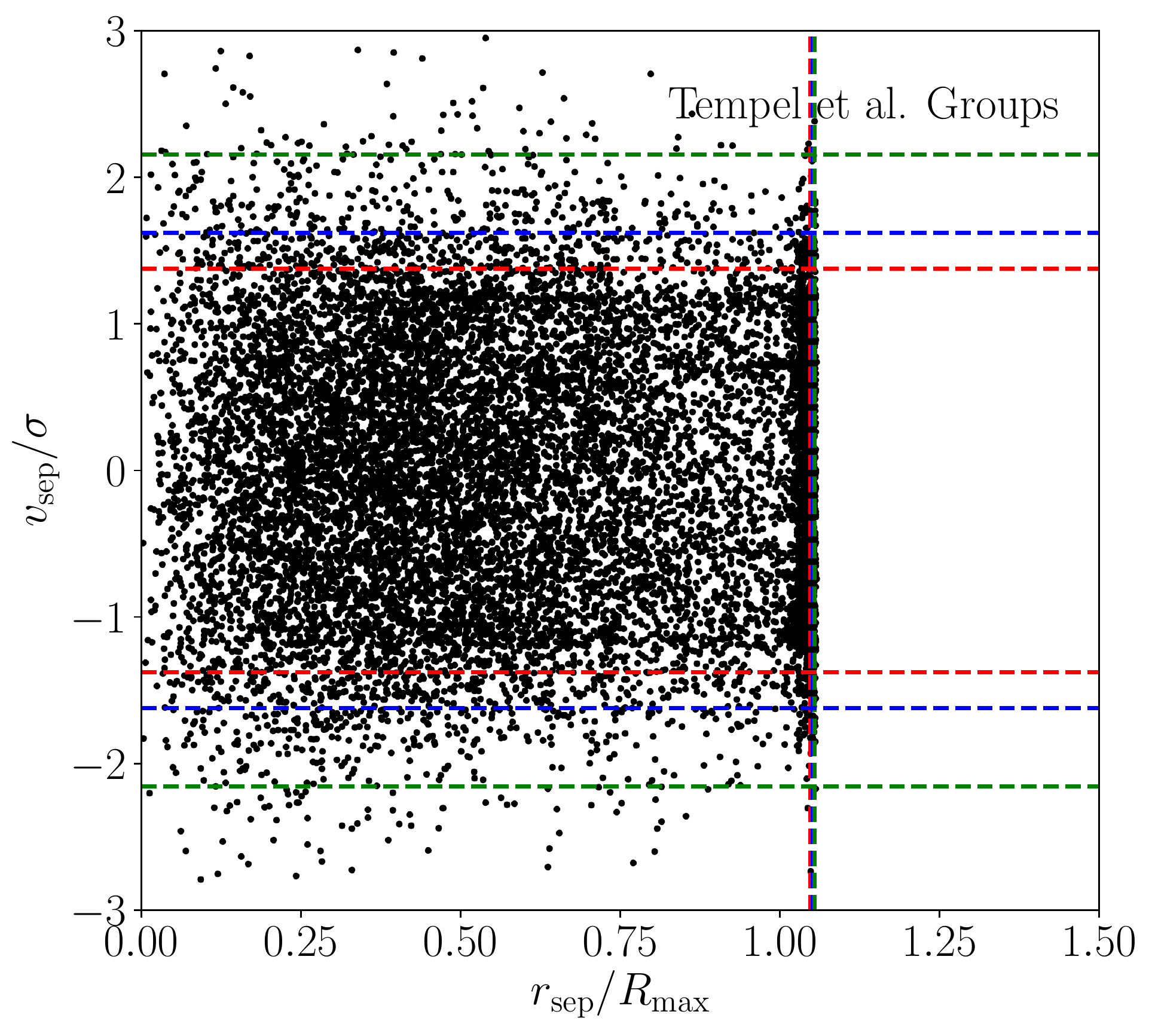}
\includegraphics[width=\columnwidth]{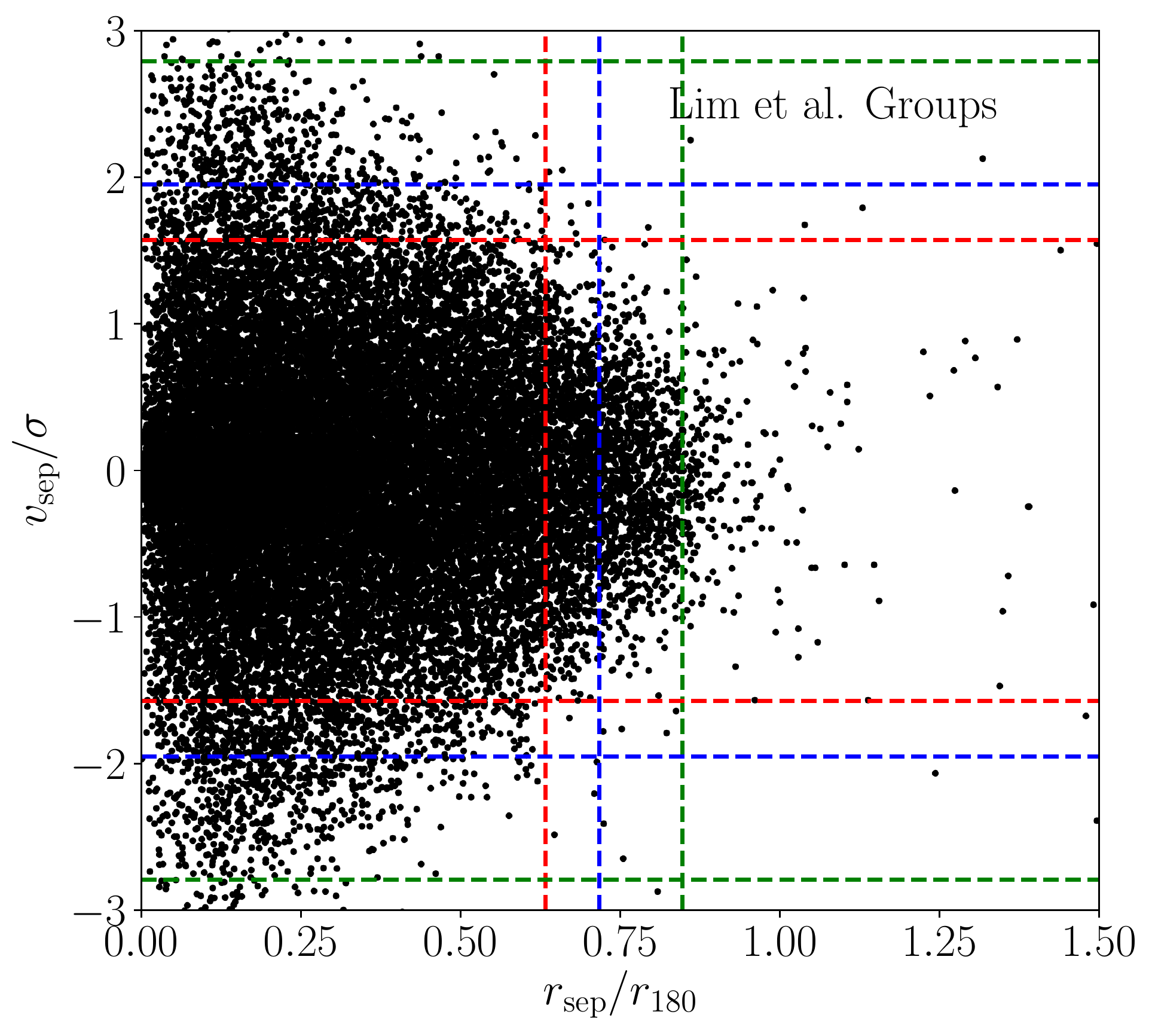}
\caption{The distribution of optical members in each group catalogue, \citetalias{Berlind+2009}, \citetalias{Yang+2012}, \citetalias{Tempel+2014} and \citetalias{Lim+2017} (top-left to bottom-right). Each member is shown as a black point in the dimensionless radial and velocity separation directions. In each case the velocity separation is scaled by the group velocity dispersion, however, as each catalogue calculates radii in different ways they are necessarily scaled differently, thus the values are not directly comparable in this dimension (except for \citetalias{Lim+2017} and \citetalias{Yang+2012}). The different radii used are described in \S\ref{sec:prox_match}. The dashed vertical lines mark the bounds that contain 90, 95, and 99 per cent of the optical members (red, blue, and green line respectively), the horizontal dashed ``tram" lines enclose the same fractions. Note these bounds were calculated independently in each dimension. In the case of \citetalias{Yang+2012} the 99 per cent lines are not shown as a smaller number of outliers shift them well beyond the range plotted.}
\label{fig:group_opt_mems}
\end{figure*}

To estimate the volumes, we sought to remain consistent with the philosophy that originally went into constructing each group catalogue, and work with the physical properties they provided.  In general, we chose values for the projected radius and line-of-sight velocity that included 90-95\% of known group members and prevented as many \hi\ detections from being matched to two groups as possible (Fig. \ref{fig:group_opt_mems}).  For \citetalias{Berlind+2009} and \citetalias{Tempel+2014} these choices (90\%) were more conservative than for \citetalias{Yang+2012} and \citetalias{Lim+2017}, and for both \citetalias{Yang+2012} and \citetalias{Lim+2017} the choices were more conservative in the velocity dimension (90\%) than in projected radius (95\%).  For matching with the \citetalias{Tempel+2014} and \citetalias{Lim+2017} catalogues we converted ALFALFA heliocentric velocities to CMB velocities to be consistent with the group catalogues.  We describe and further justify our choices below.  In the equations we retain the nomenclature used in the original works.

The \citetalias{Berlind+2009} catalogue provided the position of the group centroid, mean redshift, line-of-sight velocity dispersion, $\sigma$, and projected rms radius of the group ($R_{\perp\,\text{rms}}$).  We found that in a number of cases, the group radius or velocity dispersion was smaller than the FoF linking length, so we set an effective minimum search volume for every group.  The group volumes were then calculated from the maximum of the scaled size of the group or a fixed fraction of the linking length:
\begin{align}
    R_{\text{prox}}\, [\text{Mpc}\, h^{-1}] &= max[1.5\,R_{\perp\,\text{rms}} , 0.5\,D_{\perp}] \\
    V_{\text{prox}}\, [\text{km\, s}^{-1}] &= cz_{\odot} \pm max[1.5\,\sigma , 0.75\,D_{\parallel}].
\end{align}
$D_{\perp}$ and $D_{\parallel}$ are related to the linking lengths $b_{\perp}$ and $b_{\parallel}$ for a given sample by $D_{\perp,\parallel} = b_{\perp,\parallel}\,n_g^{-1/3} h^{-1}$ where $n_g$ is the volume density of galaxies in the sample \citepalias[Equations 3-5 of][]{Berlind+2006}.  For DR7 Mr18, $n_g = 0.03013$ (see footnote from Section 3.1).

The \citetalias{Tempel+2014} catalogue provided the position of the group centre, mean redshift corrected for the CMB, the velocity dispersion, $\sigma$, and three different radii.  These include (1) ``$\sigma_{\text{sky}}$, the rms deviation of the projected distance in the sky from the group centre'', (2) an estimate of the virial radius from the projected harmonic mean distance between galaxies, and (3) the maximum radius of the group, $R_{\text{max}}$.  In general, we found the value for $\sigma_{\text{sky}}$ corresponded to roughly one-third of the maximum radius, and the so-called virial radius was similar to $\sigma_{\text{sky}}$, but with a large scatter.  In the end we chose the largest scaled value of the maximum radius which did not result in a significant number of \hi\ detections being matched to multiple groups ($\le5$ instances for our choice).  Similar to \citetalias{Berlind+2006}, the group volumes were calculated from the maximum of this scaled size of the group or a fixed fraction of the linking length:
\begin{align}
    R_{\text{prox}}\, [\text{Mpc}\, h^{-1}] &= max[0.7\,R_{\text{max}} , 0.5\,d_{LL}] \\
    V_{\text{prox}}\, [\text{km\, s}^{-1}] &= cz_\text{cmb} \pm max([1.38\,\sigma, 0.5\,v_{LL}])
\end{align}
where $d_{LL}$ is given by Equation 2 in \citetalias{Tempel+2014}, $v_{LL} = 10 \times d_{LL}$ and both are slowly varying functions of $z_{cmb}$.

The iterative halo finders did not require a minimum group volume because the parameters reported in the catalogues were based on the estimated properties of the dark matter halos rather than the mean positions of the galaxies.
For the \citetalias{Yang+2012} catalogue, the group volumes were defined by:
\begin{align}
    R_{\text{prox}}\, [\text{Mpc}\, h^{-1}] &= 0.75\,r_{180} \\ 
    V_{\text{prox}}\, [\text{km\, s}^{-1}] &= cz_{\odot} \pm 1.6\,\sigma
\end{align}
where $r_{180}$ and $\sigma$ are given by Equations 5 and 6 in \citetalias{Yang+2007}.  The radius and velocity dispersion of the groups are calculated from the group halo mass, which is in turn estimated using a varying stellar mass-to-light ratio based on a characteristic luminosity ($L_{19.5}$).  This provides reasonable estimates for halo mass for high mass groups,
but provides significantly poorer estimates for individual low mass groups.

For the \citetalias{Lim+2017} catalogue, the group volumes were defined by:
\begin{align}
    R_{\text{prox}}\, [\text{Mpc}\, h^{-1}] &= 0.75\,r_{180} \\ 
    V_{\text{prox}}\, [\text{km\, s}^{-1}] &= cz_\text{cmb} \pm 1.57\,\sigma
\end{align}
where $r_{180}$ and $\sigma$ are given by Equation 4 in \citetalias{Lim+2017}.  As in \citetalias{Yang+2012} the radius and velocity dispersion of the groups are dependent on the halo mass, but this time use the stellar mass of all galaxies in the group and includes a correction based on the gap between the brightest central galaxy and the $n$th brightest galaxy \citep[based on conclusions from][]{Lu+2016}. This corrects the halo mass for groups which sit close to the flux limit of a shallow survey \citepalias[see Section 3.2 in][]{Lim+2017}.  The scatter in the stellar mass versus halo mass as compared to mocks is reduced across the whole group mass range and is significantly more reliable for low mass groups as compared to \citetalias{Yang+2012}.

We see from Fig. \ref{fig:group_opt_mems}, that the shape of the FoF groups are quite different in phase space than the groups determined by the iterative halo finder.  The overall shape of the FoF groups look like a cylinder projected onto phase space.  This is consistent with how FoF works: the algorithm looks for companions within a cylindrical volume centred on the seed galaxy.  
When any new galaxies are found within this volume they are added to the group and the cylindrical volume is re-centred on the new members and the search repeated.
By comparison the iterative halo finder groups look like a sphere projected onto phase space.  Iterative halo finders also use a cylindrical volume by the nature of observational limitations, however the search cylinder is centred on the weighted geometric centre of all the group members, and its size is dependent on the stellar mass of the group members. We suspect it is this more adaptive and gradually changing search volume that results in more realistically shaped groups in phase space.

A summary of the number of ALFALFA galaxies assigned to each group catalogue is given in Table \ref{tab:assign_summary}. 
After the ALFALFA galaxies have been assigned to a group their \hi \ masses are re-calculated using the redshift of the relevant group in the group catalogue and the corresponding Hubble-Lema\^{i}tre flow distance.

\begin{table*}
\centering
\caption{Summary of HI assignments to optical group catalogues}
\label{tab:assign_summary}
\begin{tabular}{lcccc}
\hline \hline
                      & Berlind et al. & Yang et al. & Tempel et al. & Lim et al. \\ \hline
Groups                & 1322           & 1579        & 3909          & 2231       \\
Minimum $cz_{\mathrm{cmb}}$ &  6750 \kms  &  3600 \kms  &  1500 \kms    &  1000 \kms \\
Maximum $cz_{\mathrm{cmb}}$ & 12500 \kms  & 15000 \kms  & 13500 \kms    & 13500 \kms \\
Optical members       & 8522           & 14280       & 16363         & 15076      \\
Limiting absolute magnitude    & -18.0 & -19.5$^\dagger$ & -18.0 & -18.7$^\dagger$\\
Minimum $r$-band luminosity    & $10^{9.07}$ \Lsol & $10^{8.21}$ \Lsol & $10^{9.07}$ \Lsol & $10^{7.10}$ \Lsol \\
Minimum \hi \ mass$^\ast$ & $10^{9.13}$ \Msol & $10^{8.59}$ \Msol & $10^{7.83}$ \Msol & $10^{7.47}$ \Msol \\
HI--optical matches   & 1891           & 2354        & 2397          & 3182       \\
HI proximity members  & 619            & 476         & 934           & 645        \\
Total HI members      & 2510           & 2830        & 3331          & 3827       \\
HI members after cuts & 1827           & 2013        & 2441          & 2865       \\ \hline
$^\dagger$ Applies only to centrals\\
$^\ast$ Assuming $W_{50} = 100$ \kms
\end{tabular}
\end{table*}

\section{\hi \ mass function calculation}
\label{sec:HIMF_calc}

The \hi \ mass function is the number density of galaxies in the Universe as a function of their \hi \ mass. Like with many astronomical distributions, the observed distribution of galaxy \hi \ masses is a highly biased representation of the intrinsic population owing to the influence of selection bias caused by survey sensitivity limits. ALFALFA is no exception to this and although the observed distribution of \hi \ masses peaks within the range $9 < \log{M_{\mathrm{H\,\textsc{i}}}/\mathrm{M_\odot}} < 10$, the lowest \hi \ mass galaxies, of which only a handful are detected, are in fact the most numerous. There are two widely used methods to correct for the survey selection bias, the $V_{\mathrm{max}}$ method \citep{Schmidt+1968} and the $V_{\mathrm{eff}}$ method \citep{Efstathiou+1988,Zwaan+2003,Martin+2010}.

Both the $V_{\mathrm{max}}$ and the $V_{\mathrm{eff}}$ methods are algorithms for estimating the volume over which galaxies of a given \hi \ mass can be detected within a given survey. By weighting detections in proportion to the inverse of these volumes the intrinsic number density of galaxies of a given \hi \ mass can be recovered. The distinction between the two methods lies in how these volumes are estimated. In the $V_{\mathrm{max}}$ method the survey completeness limit is used to determine the maximum distance that each galaxy could be placed at and still be detected. This distance is then cubed and multiplied by the area of the survey footprint to estimate the volume over which the galaxy can be detected, this volume is then referred to as the $V_{\mathrm{max}}$ for that galaxy. In the $V_{\mathrm{eff}}$ method the 2 dimensional distribution of \hi \ mass and \hi \ velocity width is split into bins and the maximum likelihood value of each bin is obtained through an iterative procedure. The velocity width dimension can then be summed over to return to HIMF. The effective weighting that is applied to each galaxy can be interpreted as an inverse volume, which is known as the effective volume or $V_{\mathrm{eff}}$.

In this work we made use of both of these methods as each has its advantages. The $V_{\mathrm{eff}}$ method is extremely robust against variations in large scale structure (LSS) within the survey volume which can cause bumps in the HIMF calculated with the $V_{\mathrm{max}}$. However, this robustness relies on the assumption that the HIMF is universal and unaffected by environment. Another property of the $V_{\mathrm{eff}}$ method is that the normalisation of the HIMF cancels in the algorithm and must be applied after the fact. While this can be a disadvantage, in this case the effective survey geometry is extremely complicated and it is much more straightforward to use the summed volume of all the groups at the end to normalise the HIMF than to consider the volume of each group within the calculation. To perform this normalisation we take the projected size of each group and assume spherical symmetry to calculate their volume. The $V_{\mathrm{eff}}$ values are then all scaled by a single factor such that the total inferred number density of galaxies matches the observed number density after the observations have been up-weighted for incompleteness \citep[refer to][Appendix A for further details]{Papastergis2013}.

The principal advantage of the $V_{\mathrm{max}}$ method is its theoretical simplicity, which permits various post hoc corrections to me made. Thus the $V_{\mathrm{max}}$ method is useful both as a comparison to the $V_{\mathrm{eff}}$ method and because it can lead to a greater understanding of the underlying causes of peculiarities in the HIMF. However, ultimately we were unable to make adequate corrections for LSS in all cases, so the following analysis relies entirely on the $V_{\mathrm{eff}}$ method.\footnote{We note here that an unresolved shortcoming of the $V_{\mathrm{eff}}$ method can lead to suppression of the first (lowest mass) few bins (discussed further in Appendix \ref{sec:vel_wid_corr}). If this effect is apparent we omit the first two bins when fitting Schechter functions to the data.} Although our analysis with the $V_{\mathrm{max}}$ method was not successful, the attempt led us to explore corrections to the method (aside from LSS corrections) which are broadly applicable to all or most existing methods for estimating the HIMF. These are discussed further in appendix \ref{sec:vmax_corrs}.

\section{Results}
\label{sec:results}

\begin{figure*}
\includegraphics[width=\columnwidth]{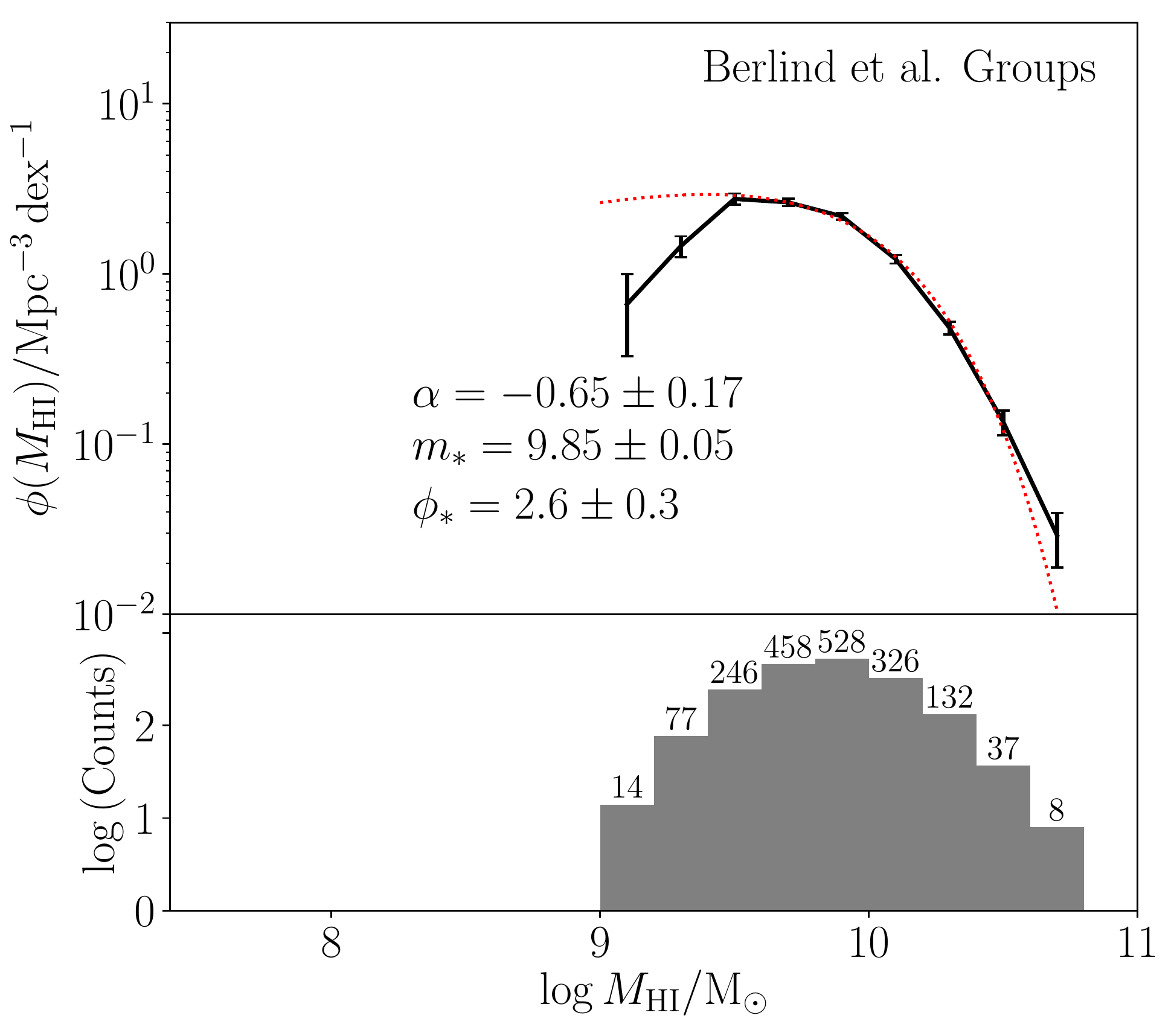}
\includegraphics[width=\columnwidth]{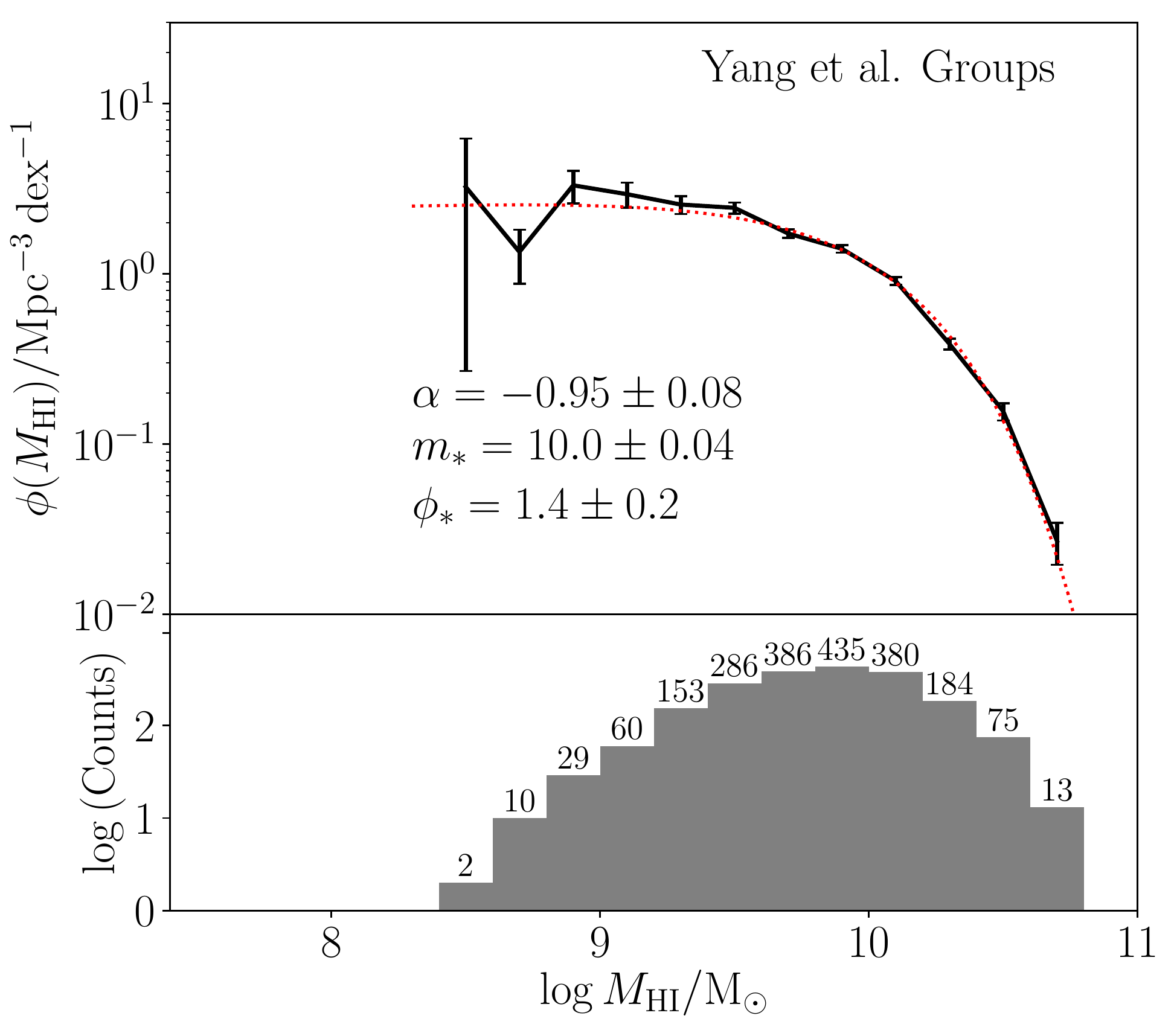}
\includegraphics[width=\columnwidth]{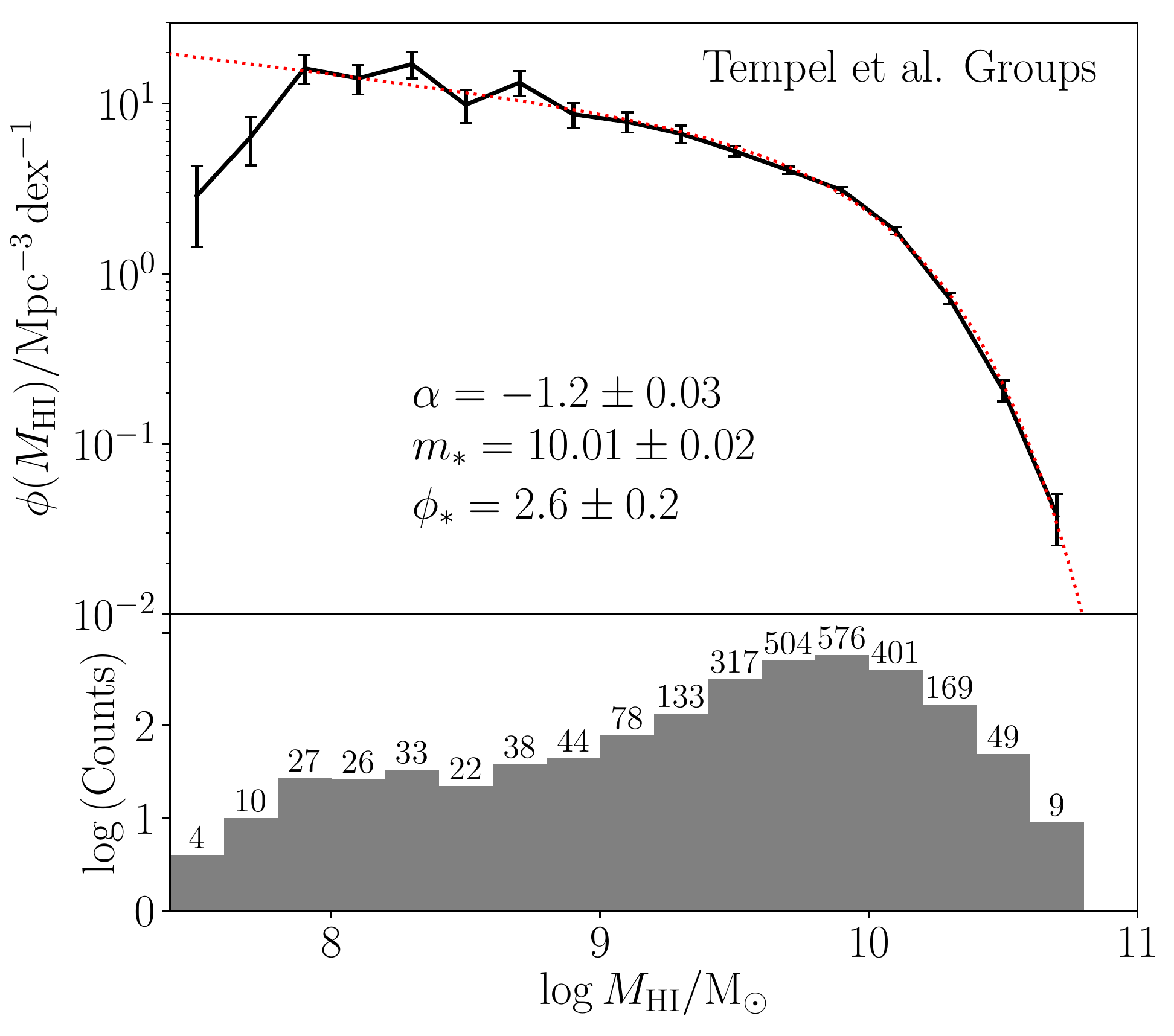}
\includegraphics[width=\columnwidth]{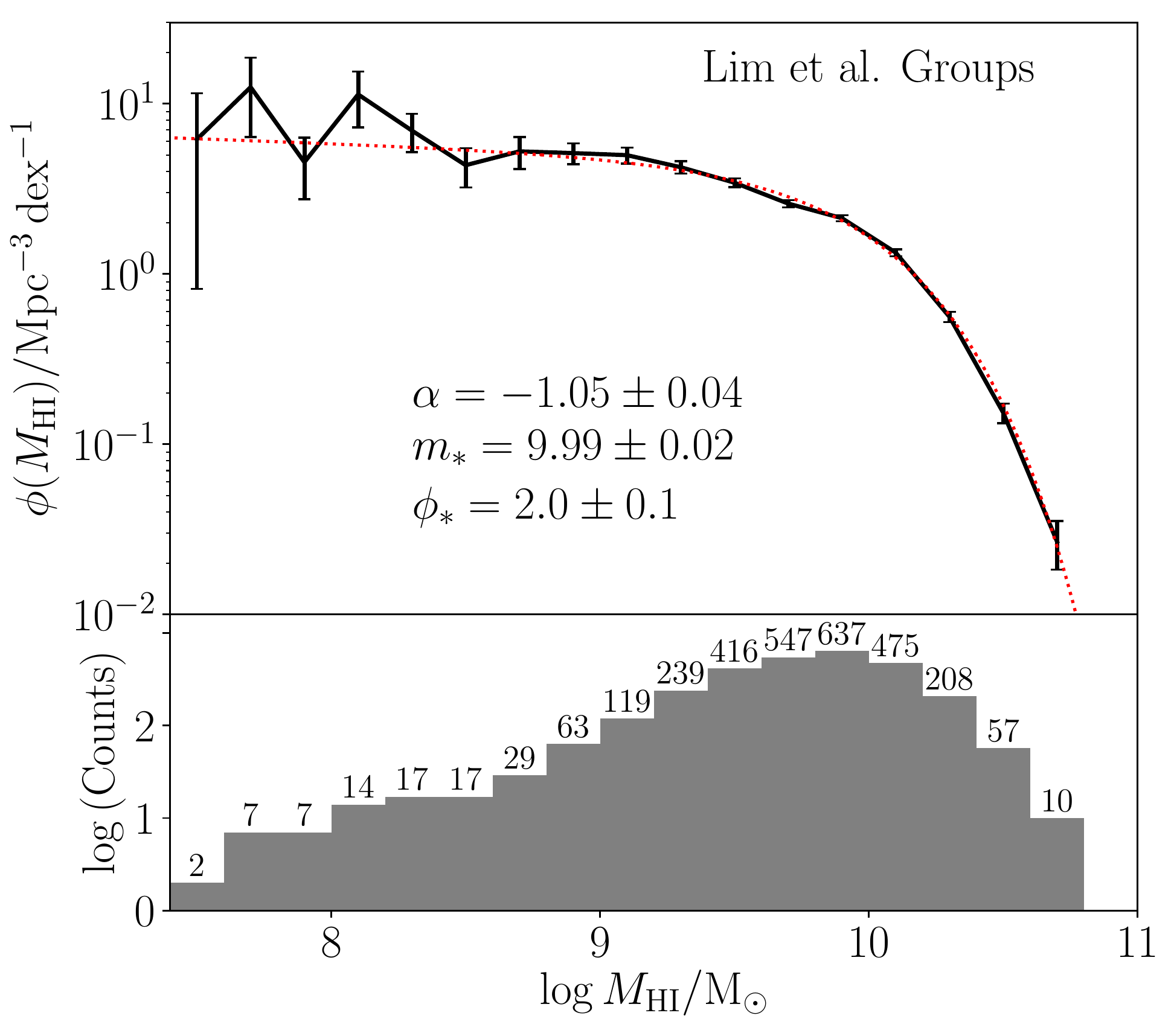}
\caption{The group galaxy HIMF (upper panels) and corresponding number counts (lower panels) of observed galaxies in each bin. From top left to bottom right the group catalogues used are \citetalias{Berlind+2009}, \citetalias{Yang+2012}, \citetalias{Tempel+2014}, and \citetalias{Lim+2017}. The HIMFs were estimated following the corrected $V_{\mathrm{eff}}$ procedure described in \S\ref{sec:HIMF_calc}. The black error bars and solid lines show the calculated number densities and associated Poisson counting errors, and the red dotted lines show the Schechter function fits. For both the \citetalias{Berlind+2009} and \citetalias{Tempel+2014} HIMF the 2 lowest mass bins are excluded from the Schechter function fit as they are likely biased (Appendix \ref{sec:vel_wid_corr}).}
\label{fig:group_HIMFs_Veff}
\end{figure*}

In this section we will first consider the HIMF of each group catalogue in turn, exploring their differences and similarities, before presenting the global results.


\subsection{The HIMFs of each group catalogue}

Fig. \ref{fig:group_HIMFs_Veff} shows the four HIMFs calculated using the $V_{\mathrm{eff}}$ method and the ALFALFA galaxies assigned to groups in each of the four catalogues. From top-left to bottom-right they are \citetalias{Berlind+2009}, \citetalias{Yang+2012}, \citetalias{Tempel+2014}, and \citetalias{Lim+2017}. The lower section of each panel displays the raw number counts of ALFALFA detections in each \hi \ mass bin, and the upper section shows the HIMF (black solid line) after these counts have been corrected by the $V_{\mathrm{eff}}$ values. The red dotted lines show the Schechter function fits to the HIMFs, the parameters of which are also quoted in each panel.

The \citetalias{Berlind+2009} group HIMF (top-left panel) shows a striking difference to the other three in that it is entirely missing sources below $\log M_{\mathrm{H\,\textsc{i}}}/\mathrm{M_{\odot}} = 9$. This is the result of the minimum distance boundary ($cz_\mathrm{cmb} > 6750$ \kms), which means ALFALFA is unable to detect lower \hi \ mass objects associated with the groups in this catalogue. This results in a poorly constrained and unrepresentative low-mass slope. Furthermore, as the two shape parameters of the Schechter function are extremely covariant, the `knee' mass should also be treated with an abundance of caution, and the quoted uncertainty is misleading due to this covariance. This HIMF also displays a clear suppression of the first two (and possibly third) bins, which we exclude from the Schechter function fit. This is also the result of the minimum distance cut, which effectively truncates part of the velocity width distribution in these lowest mass bins. The $V_{\mathrm{eff}}$ method does not suitably correct for this truncation and as a result the true abundance of galaxies in these bins is systematically underestimated (discussed further in Appendix \ref{sec:vel_wid_corr}).



Next consider the \citetalias{Yang+2012} group HIMF (top-right panel). Similarly to \citetalias{Berlind+2009}, the low-mass slope of the \citetalias{Yang+2012} HIMF is truncated by the minimum distance cut enforced in this catalogue. However, as the cut is at a considerably nearer distance ($cz_\mathrm{cmb} > 3600$ \kms) there are several well sampled bins on the low-mass slope allowing for a somewhat tighter constraint. The expected suppression of the first few bins is not seen in this case, but given the large uncertainty on those values, it is likely present but not apparent. The `knee' mass is measured to be approximately $m_\ast = 10.0$ (where $m_\ast = \log (M_\ast/\mathrm{M_\odot})$), and the low-mass slope is consistent with being flat.

In the case of the \citetalias{Tempel+2014} group HIMF (bottom-left panel) there is apparently good sampling of the low-mass slope until $\log M_{\mathrm{H\,\textsc{i}}}/\mathrm{M_{\odot}} \approx 8$, after which the bins are suppressed by the minimum distance cut ($cz_\mathrm{cmb} > 1500$ \kms) effect (and are thus excluded from the Schechter function fit). While the `knee' mass agrees with that found in the previous panel \citepalias[][top-right]{Yang+2012}, the low-mass slope is significantly steeper. In the lower section of the panel a pronounced bump is also apparent in the raw number counts. This bump can also be seen in the \citetalias{Lim+2017} panel (bottom-right), but is much less pronounced. We will return to this point in \S\ref{sec:virgo}.

Finally, consider the \citetalias{Lim+2017} group HIMF (bottom-right panel). Of the four catalogues this was assigned the most \hi \ members to its groups and extends to the lowest redshift ($cz_\mathrm{cmb} > 1000$ \kms). The `knee' mass again agrees with the previous two measurements, but the low-mass slope is marginally steeper than flat, in agreement with \citetalias{Yang+2012} (top-right panel), but in tension with \citetalias{Tempel+2014} (bottom-left panel). There is also no apparent suppression of the lowest mass bins for \citetalias{Lim+2017}, but a simple calculation explains why. The ALFALFA 50 per cent completeness limit \citep[][equations 4 and 5]{Haynes+2011} for a galaxy of a velocity width of 100 \kms \ (fairly broad for a low-mass galaxy) falls at $\log M_{\mathrm{H\,\textsc{i}}}/\mathrm{M_{\odot}} = 7.83$ and 7.47 for the minimum distance cuts in the \citetalias{Tempel+2014} and \citetalias{Lim+2017} catalogues, respectively. That means that both of the first two bin in the \citetalias{Tempel+2014} HIMF are expected to suffer from suppression, while only the first bin of the \citetalias{Lim+2017} HIMF is even partially affected.

While the three catalogues with a good sampling of \hi \ masses all agree that $m_\ast = 10.0$, slightly higher than the global ALFALFA value, the \citetalias{Yang+2012} and \citetalias{Lim+2017} groups have a flat low-mass slope, whereas \citetalias{Tempel+2014} has a steeper slope that is almost consistent (at 1$\sigma$) with that of the global ALFALFA HIMF \citep{Jones+2018}. It is unsurprising, but encouraging, that \citetalias{Yang+2012} and \citetalias{Lim+2017} agree as they employ almost the same methodology; \citetalias{Lim+2017} essentially being an extension of the \citetalias{Yang+2012} catalogue down to lower mass groups. However, the disagreement in the low-mass slope between these and \citetalias{Tempel+2014} is unexpected and must be investigated further.

\subsection{Impact of the Virgo cluster}
\label{sec:virgo}

Comparing \citetalias{Tempel+2014} and \citetalias{Lim+2017} (bottom-left and right, respectively), the difference in the low-mass galaxies can already be seen in the histograms of observed counts (before any completeness correction is made). Although these two catalogues cover the same volume (and thus the same LSS) \citetalias{Tempel+2014} has many more low-mass ALFALFA galaxies assigned to groups than \citetalias{Lim+2017}. The peak bin (just below $\log M_{\ast}/\mathrm{M_{\odot}} = 10$) has 11 per cent more galaxies in \citetalias{Lim+2017}, but in the bins either side of $\log M_{\ast}/\mathrm{M_{\odot}} = 8$ there are 53 \hi \ galaxies assigned to groups in \citetalias{Tempel+2014}, compared to only 21 in \citetalias{Lim+2017}. There is a bump in the distribution in both histograms around this mass, but clearly it is much stronger for \citetalias{Tempel+2014} than in \citetalias{Lim+2017}. A similar bump in the counts histogram of the full ALFALFA sample is seen at the same mass \citep[e.g.][]{Martin+2010,Jones+2018} and is caused by the presence of the Virgo cluster at a distance where such galaxies are just above the detection threshold, thereby enhancing the number detected. However, this is not the complete picture as the Virgo cluster is in both catalogues and the $V_\mathrm{eff}$ method corrects for this enhancement. Therefore, the difference must be in how, and how many, \hi \ galaxies are assigned to the groups in the respective catalogues.

\begin{figure*}
\includegraphics[width=\textwidth]{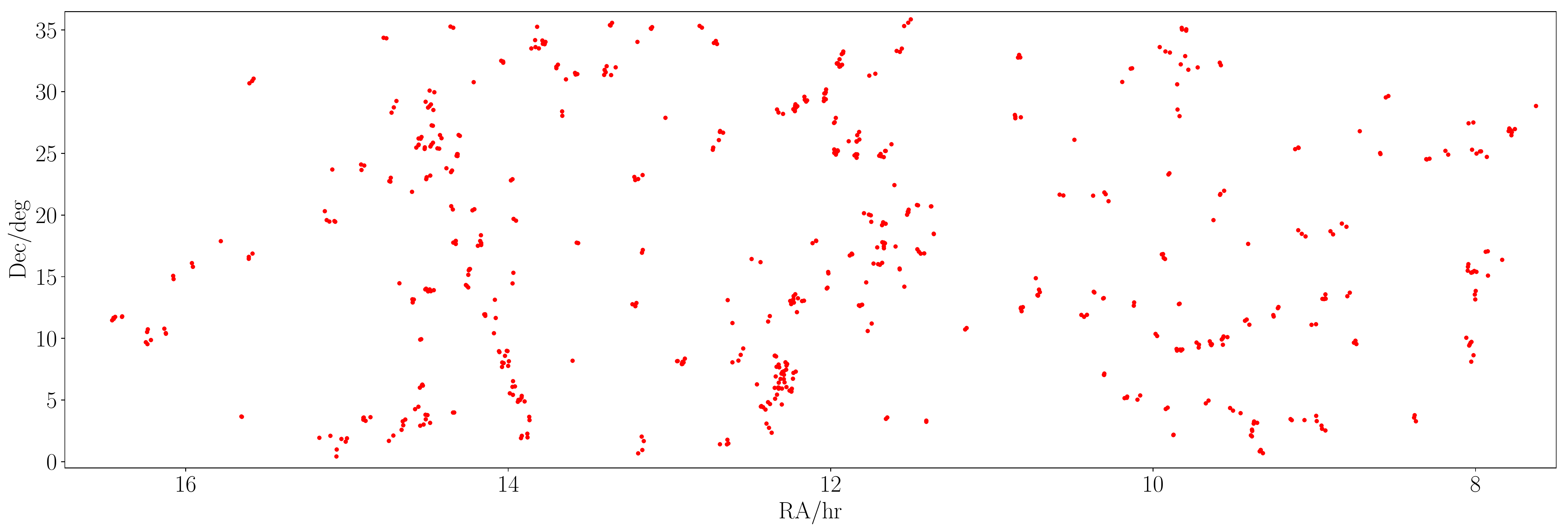}
\includegraphics[width=\textwidth]{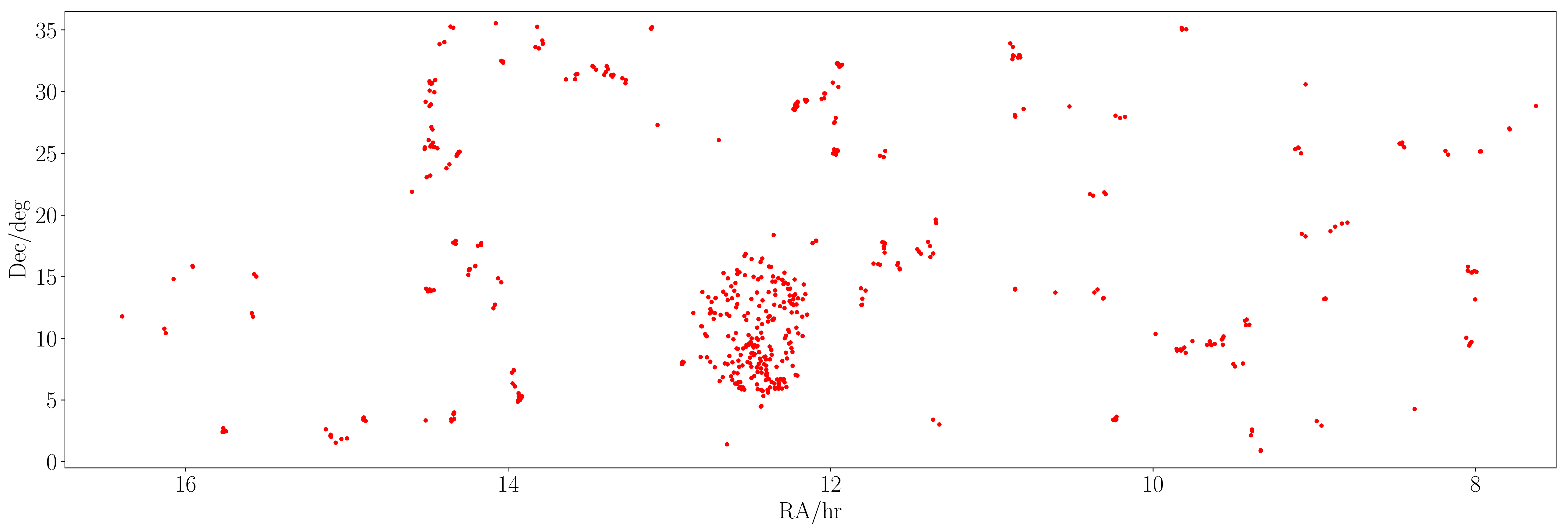}
\caption{Sky plot showing the distribution of \hi \ group members in the range $1500 < cz_\mathrm{cmb}/\mathrm{km\,s^{-1}} < 6000$ for the \citetalias{Lim+2017} (top) and \citetalias{Tempel+2014} (bottom) catalogues.}
\label{fig:group_HI_mems}
\end{figure*}

Fig. \ref{fig:group_HI_mems} shows the distribution on the sky of the \hi \ galaxies assigned to \citetalias{Lim+2017} and \citetalias{Tempel+2014} groups. The location of the Virgo cluster is immediately apparent in the \citetalias{Tempel+2014} groups (lower panel). As described in detail by \citet{Mei+2007} this region is extremely complicated and the cluster proper is surrounded by many in-falling groups. In the \citetalias{Lim+2017} catalogue the area surrounding the cluster is broken up into other, smaller groups, whilst from \citetalias{Tempel+2014} it has become an enormous single structure. The latter of these no doubt includes many galaxies which are not truly group galaxies, but just happen to be in the vicinity of Virgo. \citet{Jones+2018} found that the low-mass slope of the HIMF is particularly steep in this region of the sky, so it is unsurprising that the inclusion of many field galaxies in this region causes the slope to steepen.

\begin{figure}
\includegraphics[width=\columnwidth]{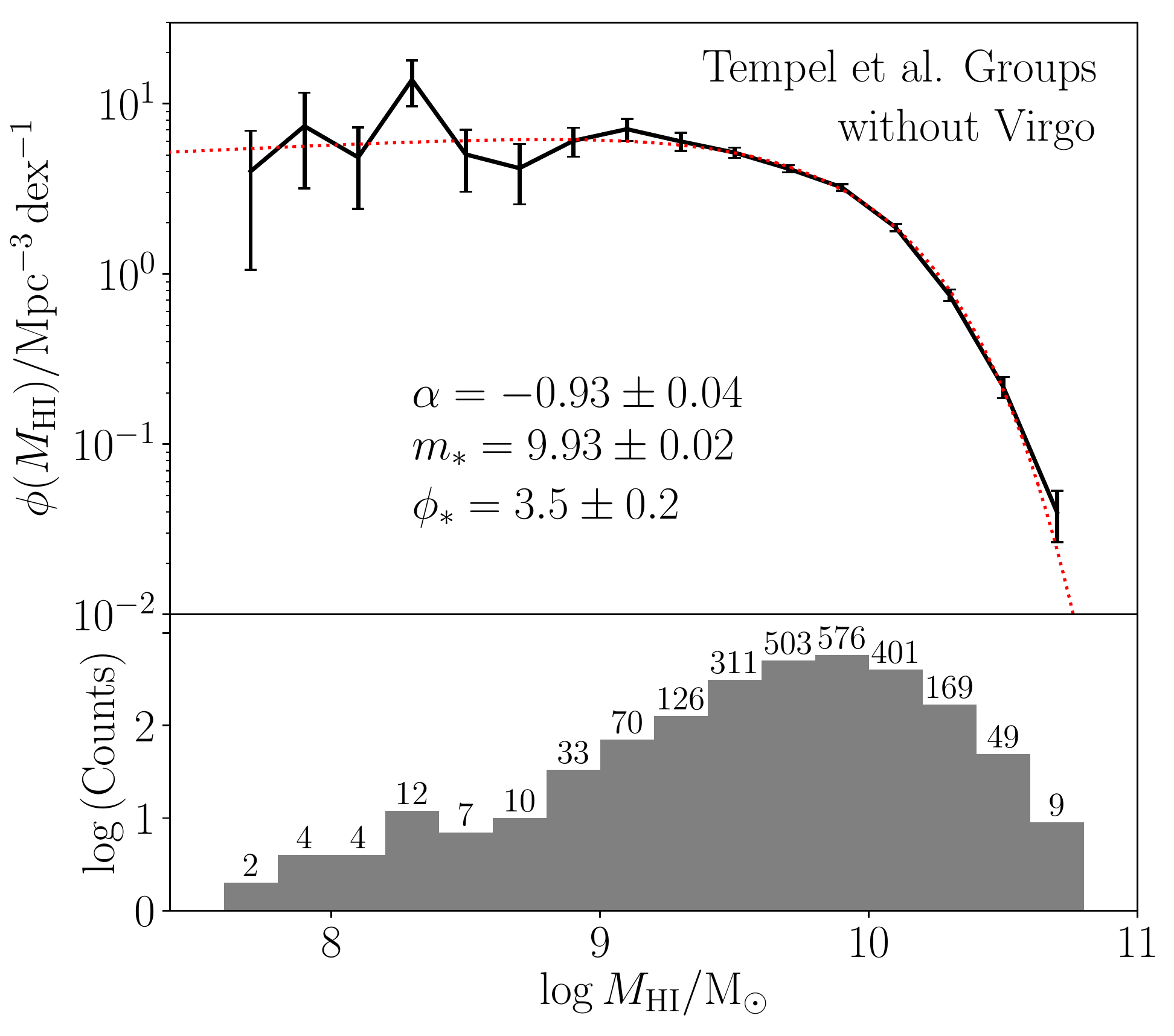}
\caption{The HIMF of the \citetalias{Tempel+2014}. group catalogue with the Virgo cluster excluded. Labelling scheme as in Figure \ref{fig:group_HIMFs_Veff}.}
\label{fig:group_HIMF_Tempel_woVirgo}
\end{figure}

Fig. \ref{fig:group_HIMF_Tempel_woVirgo} shows the group galaxy HIMF for the \citetalias{Tempel+2014} groups, calculated in the same manner as Fig. \ref{fig:group_HIMFs_Veff}, but with Virgo manually removed from the sample. This alone removes 162 \hi \ galaxies, which has the result of causing the low-mass slope to increase by approximately 0.2, making it broadly consistent with \citetalias{Lim+2017} and \citetalias{Yang+2012}. As the structure surrounding Virgo in \citetalias{Tempel+2014} is apparently single-handedly responsible for the steepness of the measured low-mass slope, and the extent of the structure cannot be considered a group, in that it is not plausible that this entire structure resides in a single parent halo, we therefore do not consider the initial result for the \citetalias{Tempel+2014} group to be valid. We will proceed with the simplistic solution of manually removing the Virgo cluster, but keep in mind that this is not an ideal solution.

\subsection{Global findings}

\begin{figure*}
\includegraphics[width=0.66\textwidth]{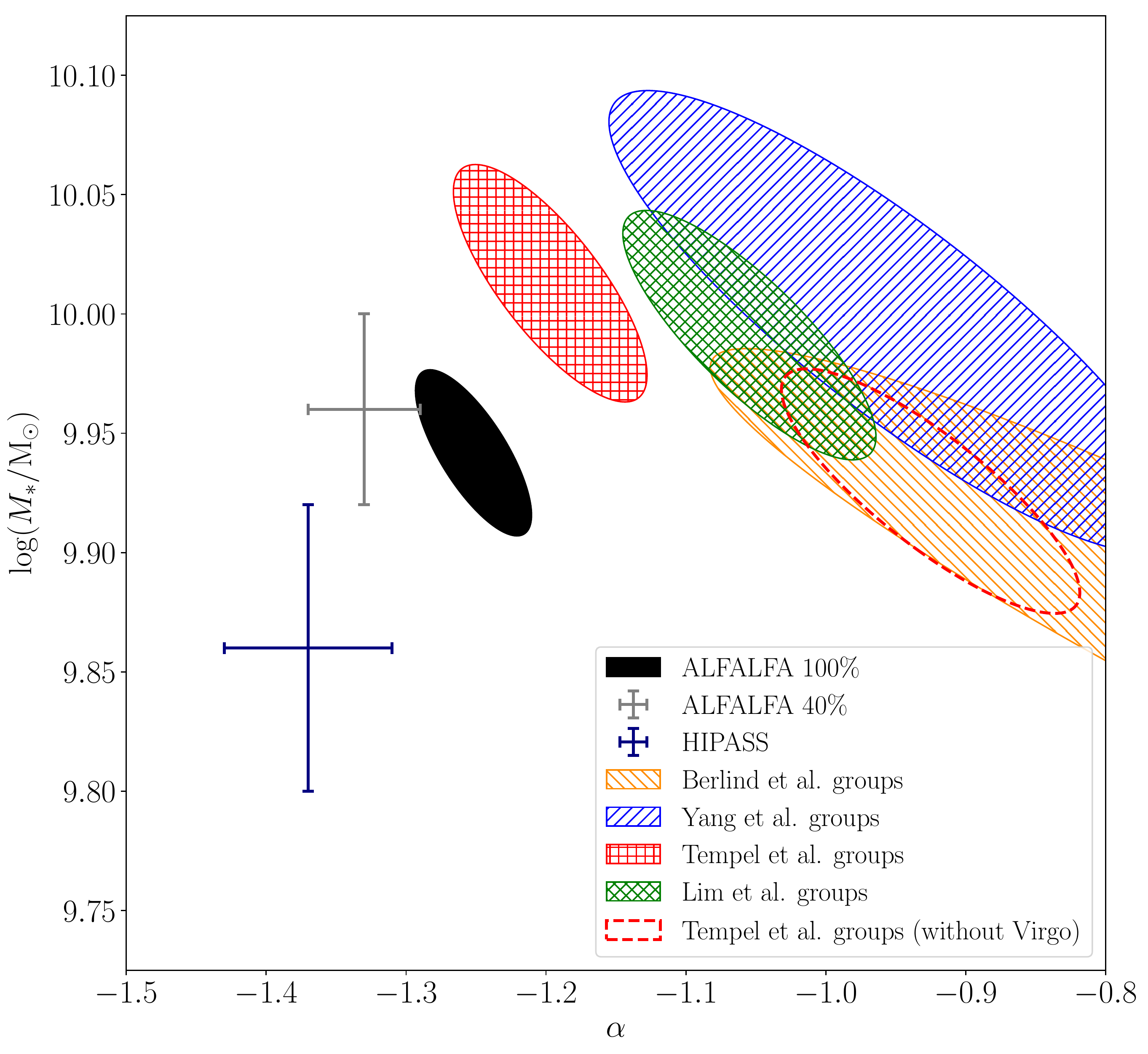}
\caption{The 2$\sigma$ error ellipses of the group HIMF Schechter function fits presented in this work and the 2$\sigma$ errorbars (and error ellipse) of the HIPASS and ALFALFA global fits. The four hatched ellipses correspond to the four groups catalogues, while the solid black ellipse indicates the location of the best fit values for the full ALFALFA sample. The grey errobars mark the location of the fit for the ALFALFA 40 per cent sample, and the dark blue errorbars that of the HIPASS sample. The thick dashed ellipse outline shows the location of the error ellipse for the \citetalias{Tempel+2014} groups without the Virgo cluster.}
\label{fig:group_HIMFs_ell}
\end{figure*}

Fig. \ref{fig:group_HIMFs_ell} shows the 2$\sigma$ error ellipses of the Schechter function shape parameters for the HIMFs shown in Figures \ref{fig:group_HIMFs_Veff} and \ref{fig:group_HIMF_Tempel_woVirgo}, as well as for the full ALFALFA sample \citep{Jones+2018}, and the 2$\sigma$ errorbars of the HIPASS \citep{Zwaan+2005} and ALFALFA 40 per cent \citep{Martin+2010} measurements. With Virgo removed from the \citetalias{Tempel+2014} groups, all four HIMFs are in agreement at the 2$\sigma$ level, preferring a significantly flatter low-mass slope than either HIPASS or ALFALFA as a whole. Although there is considerable scatter in the values of the `knee' mass, the preferred region lies marginally higher than that of the ALFALFA global value and even more so compared to HIPASS. In summary, the differences between the group catalogues notwithstanding, there appears to be a consensus that the group galaxy HIMF low-mass slope is approximately flat and that its `knee' mass is slightly higher than the global value.

\section{Discussion}
\label{sec:discuss}

In this section we briefly estimate the impact of source confusion on our measurements before proceeding to measure the HIMF in different halo mass bins, compare the group and field HIMFs, and discuss why a flattening of the low-mass slope was not seen in previous works focusing on local environment and the HIMF in ALFALFA.

\subsection{Source confusion}

A potential issue for measuring the HIMF in any of the four groups catalogues is that of source confusion. While this has been shown not to be a significant problem for ALFALFA as a whole \citep{Jones+2015}, groups represent a much denser environment than the typical ALFALFA galaxy resides in, which again raises confusion as a concern, especially as the resolution of ALFALFA is around 3.5\arcmin. In Appendix \ref{sec:confusion} we investigate the expected rate of confusion for the \citetalias{Tempel+2014} catalogue in detail and come to the conclusion that although a non-negligible fraction of the ALFALFA galaxies assigned to groups ($\sim$20 per cent) are probably, to some extent, confused, this is unlikely to have a significant impact on the their total \hi \ mass measurements. A large part of the reason for this is because, in terms of \hi \ detections, the groups are not nearly as dense as would be expected. For example, the 2865 (Table \ref{tab:assign_summary}) ALFALFA galaxies assigned to \citetalias{Lim+2017} groups are only spread amongst 1457 different groups, an average of just 2 \hi \ detections per group (excluding the 774 groups with no \hi \ detections).

\subsection{Groups of different halo mass}

\begin{figure}
\includegraphics[width=\columnwidth]{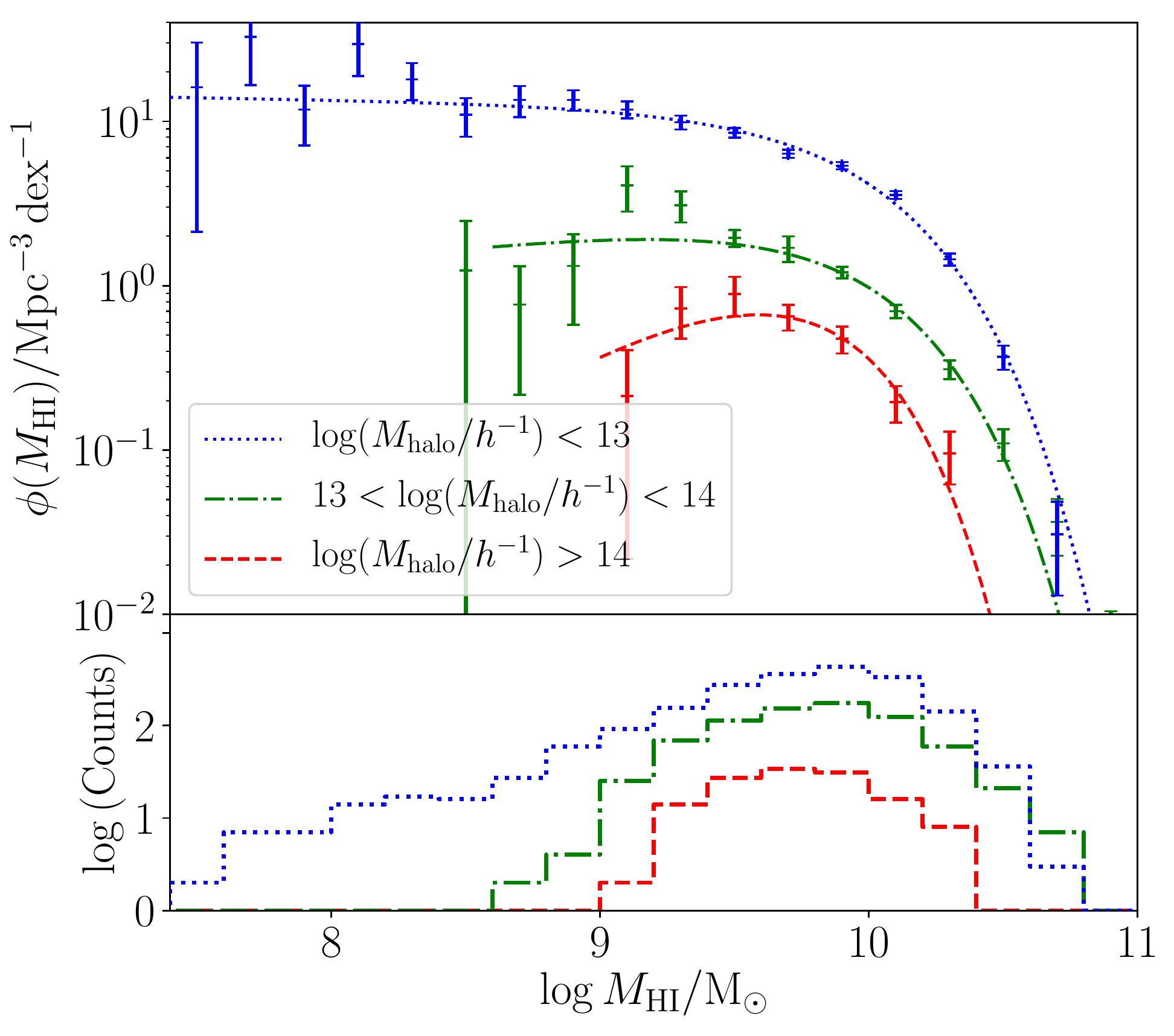}
\caption{\textit{Top panel}: The group galaxy HIMF split by group halo mass for the \citetalias{Lim+2017} groups. Schechter function fits to the data are shown with a blue dotted line (low halo mass bin), a green dash-dot line (intermediate halo mass bin), and a red dashed line (high halo mass bin). \textit{Bottom panel}: Histogram showing the raw number counts of galaxies in each halo mass bin with the same styles as above.}
\label{fig:halo_split_HIMF}
\end{figure}

\begin{table*}
\centering
\caption{Schechter function fits to group HIMFs split by halo mass}
\label{tab:halo_split_HIMF}
\begin{tabular}{cccccc}
\hline\hline
Halo mass bin           & Groups  & \hi \ members   & $\alpha$         & $m_{\ast}$      & $\phi_{\ast}/\mathrm{Mpc^{-3}\,dex^{-1}}$ \\ \hline
$\log M_{\mathrm{halo}}h/\mathrm{M_{\odot}} < 13$  &   1799 & 1938 & $-1.02 \pm 0.05$ & $9.97 \pm 0.03$ & $5.2 \pm 0.4$            \\
$13 < \log M_{\mathrm{halo}}h/\mathrm{M_{\odot}} < 14$ & 409 & 788 & $-0.82 \pm 0.16$ & $9.93 \pm 0.07$ & $1.3 \pm 0.2$            \\
$\log M_{\mathrm{halo}}h/\mathrm{M_{\odot}} > 14$  &  23  & 139 & $-0.02 \pm 0.45$ & $9.60 \pm 0.12$ & $0.79 \pm 0.11$            \\ \hline
\end{tabular}
\end{table*}

\begin{figure}
\includegraphics[width=\columnwidth]{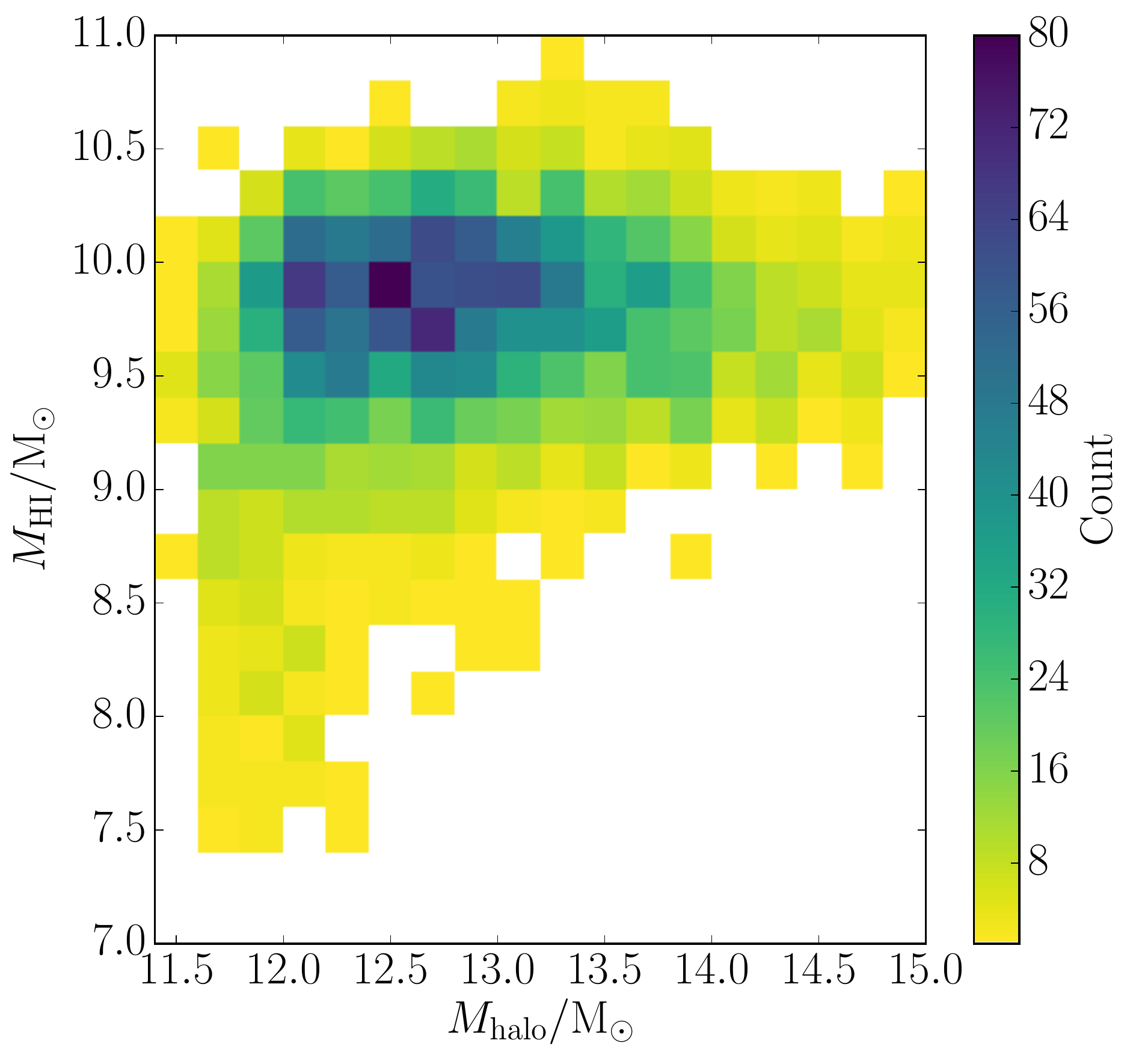}
\caption{The distribution of galaxy \hi \ masses as a function of the halo masses of their parent groups (for the \citetalias{Lim+2017} groups). This distribution has no correction for incompleteness, but highlights that the low-mass slope of the HIMF can only be constrained for low halo mass groups.}
\label{fig:MHIvMhalo}
\end{figure}

The group galaxies used to calculate the HIMF here can reside in anything from triplets to clusters, and the group galaxy HIMFs shown in Fig. \ref{fig:group_HIMFs_Veff} are in this sense averages. However, it is possible and likely that the HIMF will change its form between these extreme cases as they represent quite different environments. 

\citetalias{Lim+2017} estimated the halo masses of every group in their catalogue as part of the group finding process. We briefly describe their approach here. Their preferred estimation method was to use a central luminosity--halo mass relation with a correction factor based on the luminosity difference between the central and the 4th brightest satellite \citep{Lu+2016}. If there were fewer than 4 satellites then the faintest was used. If only the central was detected then the stellar mass--halo mass relation was used. The halo masses for all groups were then finalised by abundance matching with the halo mass function of \citet{Sheth+2001}. These steps were all repeated in every iteration of their group finding process.

To investigate the impact of halo mass we split the \citetalias{Lim+2017} catalogue into three halo mass bins based on the halo mass estimates in that work. The three mass bins were selected to approximately divide into three different regimes: 
\begin{enumerate}
\item $\log M_{\mathrm{halo}}h/\mathrm{M_{\odot}} < 13$: Below this halo mass environment is not thought to strongly affect the evolution of galaxies. 
\item $13 < \log M_{\mathrm{halo}}h/\mathrm{M_{\odot}} < 14$: This is the broad range that covers what is generally considered a group, from a structure comparable to the Local Group up to one almost 10 times more massive. In this range tidal interactions and strangulation of small galaxies may influence their properties.
\item $\log M_{\mathrm{halo}}h/\mathrm{M_{\odot}} > 14$: In this range are groups with many hundreds of members as well as some clusters. In this regime an appreciable intra-group medium is expected and ram pressure stripping should also play a role in the evolution of the galaxies.
\end{enumerate}
The \hi \ detections within each halo mass bin were used to calculate the HIMF for groups of that halo mass, following the same method as used previously. When normalising the HIMFs only groups within the same halo mass bin were considered as contibuting to the total volume and number counts. The HIMFs in these group halo mass ranges are shown in Fig. \ref{fig:halo_split_HIMF}. 

What is immediately evident is that unfortunately the low-mass slope is only well sampled for the lowest halo mass bin, which makes comparison of the slope in more massive groups difficult. Having said this, there appears to be a suggestion (Table \ref{tab:halo_split_HIMF}) that the low-mass slope actually becomes a rising slope in more massive groups, but the measurements are very uncertain. This is the result of the fact that (in comparison to optical surveys) ALFALFA is a relatively shallow survey due to the faintness of the \hi \ line. Large groups and clusters are uncommon and therefore a large volume is required to have a large sample of such objects, however, at large distances ($\gtrsim 50 \; \mathrm{Mpc}$) ALFALFA cannot detect galaxies along the majority of the low-mass slope (Fig. \ref{fig:MHIvMhalo}). This also means that the low-mass slopes of the HIMFs in Fig. \ref{fig:group_HIMFs_Veff} are also dominated by the galaxies in nearby and low halo mass groups. Therefore, the low-mass slopes should only be considered representative of such groups. To measure the low-mass slope of more massive groups would require a deeper blind survey or deep targeted observations of many groups.

Unlike the low-mass slope, the `knee' mass is well sampled in all three halo mass bins. While the $m_\ast$ values in the two lower halo mass bins are in agreement, the value for the highest bin is considerably lower. This might be indicating that in groups with $\log M_{\mathrm{halo}}h/\mathrm{M_{\odot}} > 14$ the environment is starting to impact the \hi \ content of even $L_\ast$ galaxies, causing the `knee' mass to drop. However, this result should not be over interpreted due to the weakly constrained low-mass slope and the covariance between $\alpha$ and $m_\ast$.

Finally, we consider the normalisations of these functions. We see that of the relatively small fraction of \hi \ galaxies that are in groups (Table \ref{tab:assign_summary} \& \S\ref{sec:field_himf}), the vast majority are in low mass groups, and the fraction decreases as the halo mass of the groups increases. In other words, the richness of high mass groups does not offset their rarity.

\subsection{Local environment}
\label{sec:env}

The numerous differences in the four groups catalogues used in this paper (discussed in \S\ref{sec:group_cats}) prompt the question of whether or not the galaxy environments in them will be similar. Also \citet{Jones+2016b} investigated the HIMF in different environments, but found no evidence for a flattening of the low-mass slope, even in the highest density environments. For the most part, the galaxies used to calculate the group galaxy HIMF here were included in the sample used in that study. This apparent discrepancy requires explanation.

To address these issues we calculated the projected 2nd nearest neighbour density, $\Sigma_2$, of all ALFALFA sources, following the methodology of \citet{Jones+2016b}, and a photometric measure of the tidal impact of neighbours, $Q_\mathrm{mag}$, following a similar methodology to \citet{Argudo-Fernandez+2014}. Here we will discuss the main findings of this analysis; the full details are presented in Appendix \ref{sec:app_env}.


Using these two metrics of environment, projected 2nd neighbour density and the tidal force parameter (Appendix \ref{sec:app_env}), we find that in general the ALFALFA galaxies which we have assigned to groups are concentrated in a high density, high tidal force region of the parameter space of the two metrics, relative to the ALFALFA population as a whole. This qualitative behaviour is the same for all four group catalogues, indicating that their environments (as measured by these metrics) are comparable.

\citet{Jones+2016b} separated the ALFALFA 70 per cent catalogue into quartiles of neighbour density to investigate the impact of environment on the HIMF. As the group galaxies have high values of this metric (Fig. \ref{fig:group_env}), they mainly would have been concentrated in the upper two quartiles in that analysis. Therefore, it is surprising that that work did not find any flattening of the low-mass slope in the higher neighbour density quartiles, given that we have measured a flat slope for group galaxies. However, this can be explained by the combination of two factors. Firstly, environment metrics generally have a large scatter and so some field galaxies also fall in the region of the parameter space where the group galaxies are concentrated. Secondly, ALFALFA is comprised mainly of field galaxies. Together these mean that there is no range of the neighbour density metric where group galaxies dominate the population (and thus the HIMF). In other words, neighbour density is not sufficient to reliably separate (\hi-selected) field and group galaxies. This explains the apparent tension between our present results and those of \citet{Jones+2016b}, and also suggests that it may be advantageous to combine multiple environment metrics, to mitigate their large scatter, when studying environmental effects.

While the 2nd nearest neighbour density metric used by \citet{Jones+2016b} may not be able to reliably separate group from field galaxies, that work did demonstrate that different quartiles of nearest neighbour density trace different density structures (their fig. 4 \& 8). With the lowest $\Sigma$ objects being approximately uniformly distributed, or even avoiding groups and clusters, and the highest $\Sigma$ objects clumping around filaments, groups, and clusters. However, although the higher density quartiles are still dominated by field objects, it is important to remember that ``field" covers a wide range of environments, from objects on the outskirts of groups or filaments, to those in voids. Taken together these two results imply that the flattening of the low-mass slope of the HIMF only occurs for galaxies within the bound volume of a group, and not within the larger scale overdensities that often surround them, in other words, only for satellite objects.

\subsection{Field HIMF}
\label{sec:field_himf}

\begin{figure}
\includegraphics[width=\columnwidth]{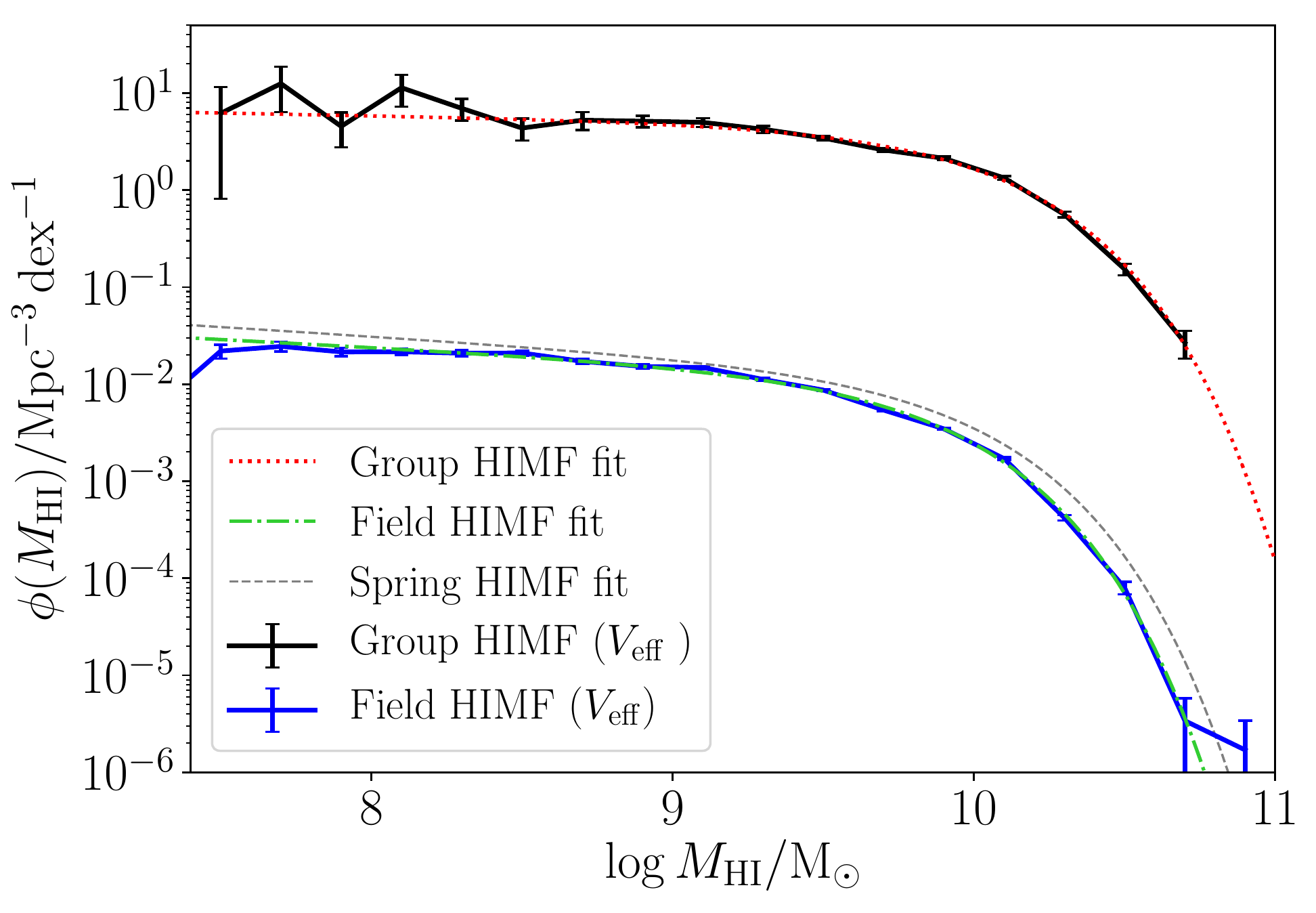}
\includegraphics[width=\columnwidth]{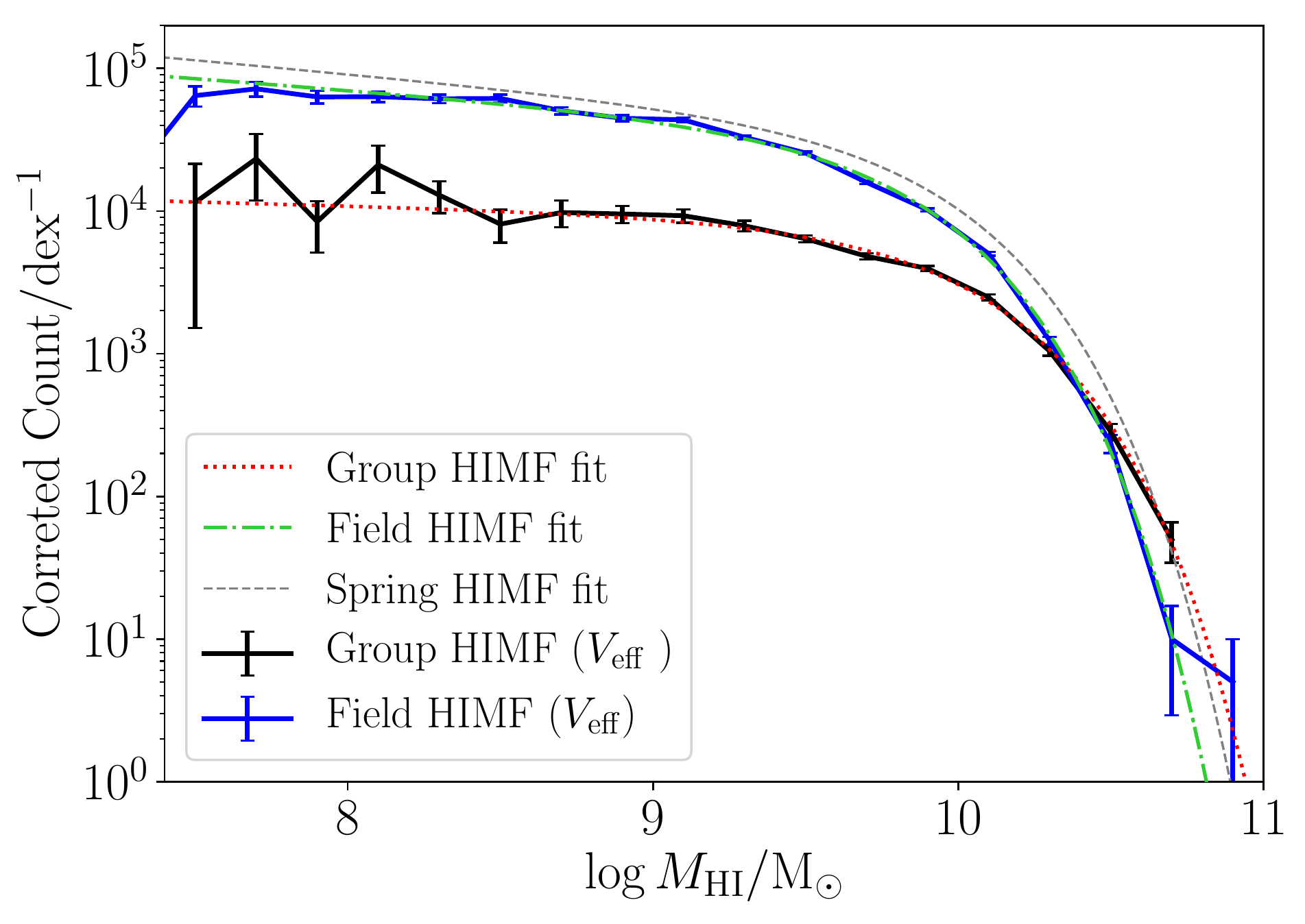}
\caption{The field HIMF (blue solid line and error bars). Calculated from the ALFALFA galaxies in the Spring sky with the \citetalias{Lim+2017} groups removed. The green dash-dot line is the Schechter function fit. The group galaxy HIMF from Figure \ref{fig:group_HIMFs_Veff} is shown for comparison (black solid line and error bars) along with its fit (red dotted line). The thin grey dashed line is the best Schechter function fit to the full ALFALFA HIMF for the Spring sky (with the same distance limits as the \citetalias{Lim+2017} sample). \textit{Top}: The HIMFs plotted following the standard convention, number density per dex.  \textit{Bottom}: The HIMFs multiplied by the volume they correspond to, giving the total corrected number counts per dex. This gives a fairer impression of how groups and the field contribute to the global galaxy HIMF.}
\label{fig:field_HIMF}
\end{figure}

The ALFALFA sample is predominantly a field sample of galaxies as HI-selected galaxies generally reside in low density environments \citep{Papastergis+2013}. However, the global ALFALFA HIMF \citep{Martin+2010,Jones+2018} is not, strictly speaking, the HIMF of field galaxies as all of the galaxies we have assigned to groups are contained within the full sample. Hence, to make a measurement of the field HIMF we must remove the galaxies we have assigned to groups. 

To do this we took the ALFALFA Spring sky region of the survey, which covers approximately the same sky area as the groups catalogues. All the ALFALFA galaxies which were assigned to a \citetalias{Lim+2017} group were removed and the sample was cut to have the same redshift limits as that catalogue ($1000 < cz_{\mathrm{cmb}}/\mathrm{km\,s^{-1}} < 13500$). This left a total of 10256 galaxies above the completeness limit. Comparison with the values in Table \ref{tab:assign_summary} indicates that on average 22 per cent of ALFALFA galaxies reside in groups (as defined by \citetalias{Lim+2017}).\footnote{This is the observed percentage and is therefore biased towards more massive galaxies. The value fairly including all \hi-bearing galaxies would be considerably lower.} 
We then proceeded with the standard $V_{\mathrm{eff}}$ method to calculate the field galaxy HIMF, which is shown by the thick, solid, blue line in Fig. \ref{fig:field_HIMF}. The Schechter function fit parameters for this field HIMF are: $\alpha = -1.16 \pm 0.02$, $m_\ast = 9.81 \pm 0.02$, and $\phi_\ast = (5.4 \pm 0.3) \times 10^{-3} \; \mathrm{Mpc^{-1}\,dex^{-1}}$. For comparison we also show the Schechter function fit to the HIMF of all the ALFALFA galaxies in the Spring sky within the same redshift limits ($\alpha = -1.19 \pm 0.02$, $m_\ast = 9.89 \pm 0.02$, and $\phi_\ast = (5.9 \pm 0.4) \times 10^{-3} \; \mathrm{Mpc^{-1}\,dex^{-1}}$), and the \citetalias{Lim+2017} group galaxy HIMF, as in Fig. \ref{fig:group_HIMFs_Veff}. Note that the former is not the same as the Spring sky HIMF presented in \citet{Jones+2018} because of the redshift limits used. The $V_{\mathrm{eff}}$ method does not completely correct for the velocity width dependence of the detection limit for the lowest mass sources, however, over the mass range covered by the group HIMF this does not appear to have led to substantial suppression of the bins (although this does occur to the left of the range plotted).

There are two panels of Fig. \ref{fig:field_HIMF} as a direct comparison of the group and field HIMFs is somewhat subjective, owing to the difference in the physical volumes for which they are relevant. In the upper panel the HIMF is plotted in the normal fashion, that is, as a number density (per dex). In this case the group HIMF is much higher than the field HIMF, but this is simply because groups are dense collections of galaxies compared to the field, so this result comes as no surprise. 

If instead the HIMF is multiplied by the total volume to which it applies then the vertical axis instead becomes a corrected number of galaxies, of a given mass, that are within that volume. For the group HIMF this volume is the sum of the volumes of all the groups (including those with no \hi \ detections), whereas for the field HIMF it is the entire survey volume (in the Spring sky, within the redshift range) minus the volume of the groups, although this is a negligible correction. When weighted in this manner we see that actually the group HIMF makes a minimal contribution to the overall HIMF over most of the mass range. However, at the high mass end the `knee' of the field HIMF falls at a lower mass and the galaxies in groups become the dominant contribution (Fig. \ref{fig:field_HIMF}, lower panel).

\section{Conclusions}
\label{sec:conclude}

We have used four popular SDSS galaxy groups catalogues \citep{Berlind+2009,Yang+2012,Tempel+2014,Lim+2017} to identify ALFALFA \hi-detected galaxies in groups and to measure the HIMF for group galaxies using the $V_\mathrm{eff}$ method. Differences in how each group catalogue was constructed lead to differences between the group galaxy HIMFs derived for each, thus there is no single group galaxy HIMF. While the \citetalias{Lim+2017} and \citetalias{Yang+2012} HIMFs are very similar, the \citetalias{Tempel+2014} catalogue includes more low-mass, \hi-rich galaxies and thus a different low-mass slope, whereas the \citetalias{Berlind+2009} catalogue is cut at a sufficiently high redshift that no low-mass galaxies are included. However, in general we find that the low-mass slope is approximately flat, significantly different from the global ALFALFA HIMF, but in agreement with studies of individual groups, and that the `knee' mass is slight higher than that of the global ALFALFA HIMF. 

The group galaxies were removed from the ALFALFA source catalogue in the Northern Spring sky and the field HIMF was calculated with the remaining galaxies, following the $V_\mathrm{eff}$ method. We find that the field HIMF is almost equivalent to the global ALFALFA HIMF in the same region of the sky, indicating that group galaxies make only a small contribution to the global HIMF, or equivalently, that the vast majority of \hi-selected galaxies are not satellites. This is most true for low-mass galaxies, as the relative contribution of groups increases with increasing \hi \ mass, with group galaxies actually becoming the dominant contribution beyond the `knee'.

We estimated the environment of all galaxies in ALFALFA, using the second nearest SDSS spectroscopic neighbour density and a photometric measure of the tidal influence of neighbours. Groups are concentrated in a higher neighbour density and higher tidal tidal influence region of the parameter space, but still overlap with the general ALFALFA population. Due to the far greater number of field galaxies in ALFALFA than group galaxies, this means that there is no region of this environment parameter space where group galaxies are the dominant population. This likely explains why previous studies of the environmental dependence of the HIMF with ALFALFA have not found significant evidence of a flattening of the low-mass slope, despite it being flat in groups.

Finally, we attempted to divide group galaxies into bins of their host halo masses. However, we find that due to the rarity of high halo mass groups, an insufficient number are nearby, where existing \hi \ surveys can detect low-mass galaxies, thus we were unable to confidently make a measurement of the low-mass slope for intermediate or high mass groups. Such an analysis will require a significantly deeper blind \hi \ survey such as the upcoming DINGO and MIGHTEE-HI surveys that will be carried out with Square Kilometre Array precursor telescopes. This is also a reminder that the HIMFs calculated for group galaxies in this work contain many different groups, and that the shape of the HIMF is dominated by galaxies in low halo mass groups as these are the most numerous.

\section*{Acknowledgements}
We acknowledge the work of the entire ALFALFA team for observing, flagging and performing signal extraction. We thank the anonymous referee for their suggestions which helped to improve this paper. MGJ is supported by a Juan de la Cierva formaci\'{o}n fellowship (FJCI-2016-29685). MGJ and LVM also acknowledge support from the grants AYA2015-65973-C3-1-R and RTI2018-096228-B-C31 (MINECO/FEDER, UE). The research of KMH is supported by the European Research Council under the European Union's Seventh Framework Programme (FP/2007-2013)/ERC Grant Agreement nr.~291531. EAKA is supported by the WISE research programme, which is financed by the Netherlands Organisation for Scientific Research (NWO). This work has been supported by the Spanish Science Ministry ``Centro de Excelencia Severo Ochoa'' program under grant SEV-2017-0709. This research was supported by the Munich Institute for Astro- and Particle Physics (MIAPP) which is funded by the Deutsche Forschungsgemeinschaft (DFG, German Research Foundation) under Germany's Excellence Strategy --- EXC-2094 --- 390783311.



\bibliographystyle{mnras}
\bibliography{refs}



\appendix

\section{Corrections for HIMF calculation methods}
\label{sec:vmax_corrs}

Although the final results of this work used the $V_{\mathrm{eff}}$ method (\S\ref{sec:HIMF_calc}), we experimented extensively with applying corrections to the $V_{\mathrm{max}}$ method for various biases stemming from an inadequate accounting of the velocity width dependence and the location of the bin edges. Ultimately we were not successful in creating suitable LSS corrections for all group catalogues and therefore focused on the $V_{\mathrm{eff}}$ method, which is very robust to LSS variations. 

Groups are a specific intermediate environment, which may not be representative of the global trends in LSS. We therefore attempted to use the groups themselves to make this correction. However, this resulted in another problem of choosing how to weight the importance of groups of different mass or memberships. If left unweighted the low mass groups dominated the correction, but the range in the other physical parameters (e.g. mass and radius) is so large that weighting by these would result in the opposite problem. We were unable to find an adequate resolution and as a result the corrected $V_{\mathrm{max}}$ HIMFs we calculated displayed un-physical bumps and curves. Therefore, these attempted were abandoned in favour of the $V_{\mathrm{eff}}$ method which does not require an explicit correction for LSS.

However, the other corrections (or equivalents) are relevant to all or most other methods used to estimate the HIMF, thus a brief description of these corrections is worthwhile.

\subsection{Velocity width correction}
\label{sec:vel_wid_corr}

The ALFALFA completeness limit is a function of both integrated flux (or equivalently, \hi \ mass at a given distance) and the velocity width of the source, $W_{50}$, as described in \citet{Giovanelli+2005} \& \citet{Haynes+2011}. This is a general property of any blind \hi \ survey as the broader the emission line, the more frequency channels contribute noise, and thus the lower the integrated signal-to-noise ratio (for a fixed total emission flux). At a fixed \hi \ mass the distribution of $\log W_{50}$ can be approximated by a Gumbel distribution with the long tail extending towards low velocity widths \citep[e.g.][]{Martin+2010,Jones+2015}. The distribution deviates slightly from a normal Gumbel distribution as it is truncated at low velocity widths due to a combination of the velocity resolution of the survey and the onset of turbulence physically preventing lower velocity widths, even in perfectly face-on galaxies. For ALFALFA we take this value to be 15 \kms, which corresponds to emission spanning three channels. We will refer to this distribution of velocity widths at fixed \hi \ mass as the mass-conditional velocity width function, or $p(w|m)$.

Typically in the $V_{\mathrm{max}}$ method the flux and $W_{50}$ measurements of each source would be used to estimate $D_{\mathrm{max}}$, the maximum distance at which it could be detected, i.e. the distance at which its flux would fall on the completeness limit, given its velocity width. The process is repeated for each source and then the inverse of all the $V_{\mathrm{max}}$ values are summed in their corresponding mass bins to produce the binned HIMF. We will refer to the $V_{\mathrm{max}}$ estimates from this approach as $V_{\mathrm{max-std}}$, or the standard $V_{\mathrm{max}}$ estimates.

While in ideal circumstances this approach does produce a reasonable approximation to the HIMF, it should be noted that this in fact does not fully account for the velocity width dependence of the completeness limit. Instead what is happening is that within each mass bin the over- and under-estimates of the galaxy number density, resulting from the $V_{\mathrm{max-std}}$ estimates from broad and narrow velocity width galaxies, mostly cancel out. However, in the case where there is a minimum distance cut (that is, when the survey volume does not extend all the way to the observer) the broadest velocity width sources in the lowest few mass bins may not be detectable anywhere in the survey volume. Thus, the resulting HIMF will systematically underestimate the number density in those bins, potentially producing an artificial flattening of the low-mass slope. As the variance of the mass-conditional velocity width function decreases at lower \hi \ masses \citep{Martin+2010,Jones+2015} this effect generally becomes less and less pronounced for progressively lower mass sources, or equivalently for progressively nearer minimum distance cuts.

It should also be noted that although the $V_{\mathrm{eff}}$ method does fully account for the 2D shape of the completeness limit by maximising the likelihood of the galaxy number density across a 2D grid of bins in both \hi \ mass and velocity width \citep[e.g.][]{Zwaan+2003,Zwaan+2005,Martin+2010,Papastergis2013}, it is only able to find a solution in bins where there are data. Therefore, the $V_{\mathrm{eff}}$ method is also not immune to this source of bias and, more generally, nor is any method that does not include a prior estimate of the mass-conditional velocity width function, $p(w|m)$.\footnote{The recently developed modified maximum likelihood method \citep{Obreschkow+2018} does have the capability to account for this source of bias, but a representative functional form for the 2D mass-width function would still need to be selected a priori. It is not possible to fully correct for this bias without applying some form of prior for the distribution of velocity widths at a given mass.} 

To make the correction the first step is to calculate the $D_{\mathrm{max}}$ values for each source using the same minimum value of $W_{50}$ (15 \kms) for all sources. The corresponding $V_{\mathrm{max}}$ values will be referred to as $V_{\mathrm{max-mass}}$, or the mass-only $V_{\mathrm{max}}$ estimates. This means that each source is counted as if the survey can detect sources of that mass, with any velocity width, over the entire volume out to $D_{\mathrm{max}}$. This is of course false because, at any given mass, the fraction of the mass-conditional velocity width distribution that is above the completeness limit is a function of distance as the observed flux changes with distance (at fixed \hi \ mass). Therefore, to calculate the width-corrected maximum volume ($V_{\mathrm{max-wc}}$) it is necessary to integrate over distance with each infinitesimal volume scaled by the fraction of $p(w|m)$ that lies above the completeness limit at that distance.
\begin{equation}
\label{eqn:vmax_wc_gen}
V_{\mathrm{max-wc}}(m) = \int_{D_{\mathrm{min}}}^{D_{\mathrm{max}}(m)} \Omega(D)  D^{2} \int_{w_{\mathrm{min}}}^{w_{\mathrm{max}}(m,D)} p(w|m) \;\mathrm{d}w\,\mathrm{d}D,
\end{equation}
where $\Omega$ is the solid angle of the survey footprint, $D_{\mathrm{min}}$ is the inner distance boundary of the entire survey volume, $w_{\mathrm{min}} = \log (15)$, $p(w|m)$ is a Gumbel distribution fit to the mass-conditional velocity width function as defined in equations C1-C3 of \citet{Jones+2015}, and $w_{\mathrm{max}}(m,D)$ is the velocity width for which a galaxy of log \hi \ mass $m$, at a distance $D$, falls on the completeness limit line \citep[obtained from equations 4 and 5 of][]{Haynes+2011}, that is, the maximum velocity width at which sources of that mass can be detected with high completeness. In general the survey footprint can vary with redshift, for example with very broad band receivers, or for surveys using multiple bands, the primary beam diameter can change considerably over the full bandwidth, or in the case of persistent RFI there may be coverage gaps at certain frequencies.

Unfortunately it is less straightforward to correct for this bias in the $V_{\mathrm{eff}}$ method. Therefore, rather than making a correction, when it was apparent (as in the cases of the \citetalias{Berlind+2009} and \citetalias{Tempel+2014} groups) the first two bins were ignored when fitting the Schechter function. This is a non-ideal solution and the existing methods need to be modified to account for this effect, especially as several future surveys will seek to measure the HIMF in redshift bins and may not have sufficient numbers of sources to be able to discard the two lowest mass bins.

\subsection{Alignment of the 1st bin}

Another effect which can cause suppression of the lowest mass bin of the HIMF is whether or not the (left-most) bin edge aligns well with the minimum detectable \hi \ mass. This effect is most noticeable when there are 10s or 100s of galaxies in the first bin, as otherwise the suppression can easily be hidden by the large Poisson noise. If the minimum detectable \hi \ mass lies in the middle of a bin then only part of the bin is actually accessible to the survey, thus the estimated counts for that bin will be an underestimate. This effect is only relevant in cases where there is a minimum distance cut, as otherwise the minimum detectable \hi \ mass is technically zero, so the bin alignment is not important. 

This effect can be avoided by using a method which does not bin the HIMF \citep[see][]{Obreschkow+2018} or by aligning the mass binning scheme with the minimum detectable mass. However, as the detection limit also depends on the velocity width, the minimum detectable mass is not fixed solely by the minimum distance cut. If the velocity width correction is applied as described above then the appropriate minimum mass to use would be that associated with the narrowest velocity width considered.

A downside of any binning method is that the choice of binning will inevitably impact the values of the final fit of the assumed functional form, in this case a Schechter function. As we want to compare the results between four different group catalogues (with different redshift limits) we do not wish to alter the binning scheme between the catalogues. Therefore, we do not correct for this effect. However, in the \citetalias{Berlind+2009} and \citetalias{Tempel+2014} catalogues we did not fit the first two bins (due to the suppression from the velocity width effect discussed above), so it was indirectly addressed anyway.

\section{Local environment}
\label{sec:app_env}

\begin{figure*}
\includegraphics[width=\columnwidth]{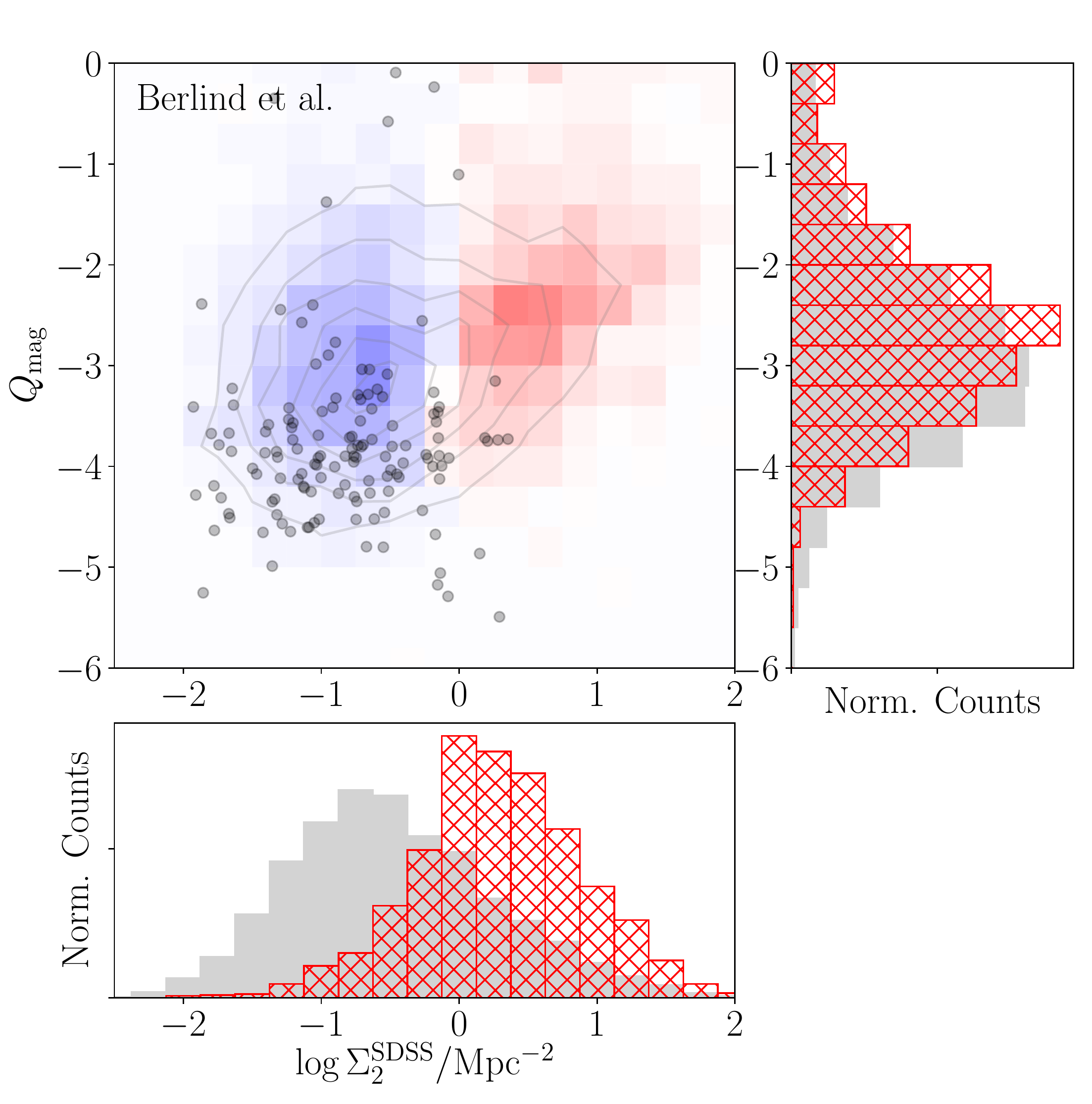}
\includegraphics[width=\columnwidth]{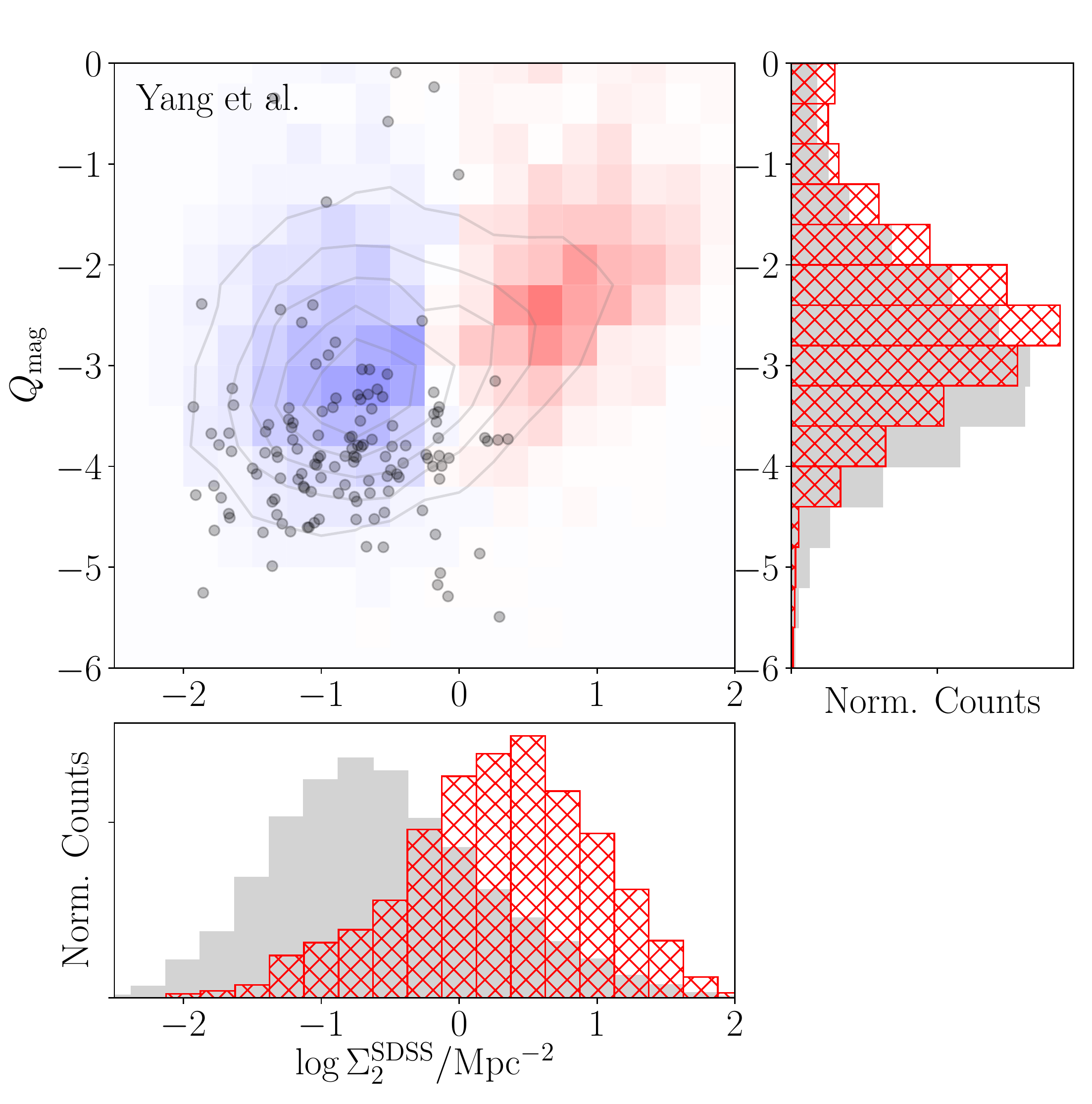}
\includegraphics[width=\columnwidth]{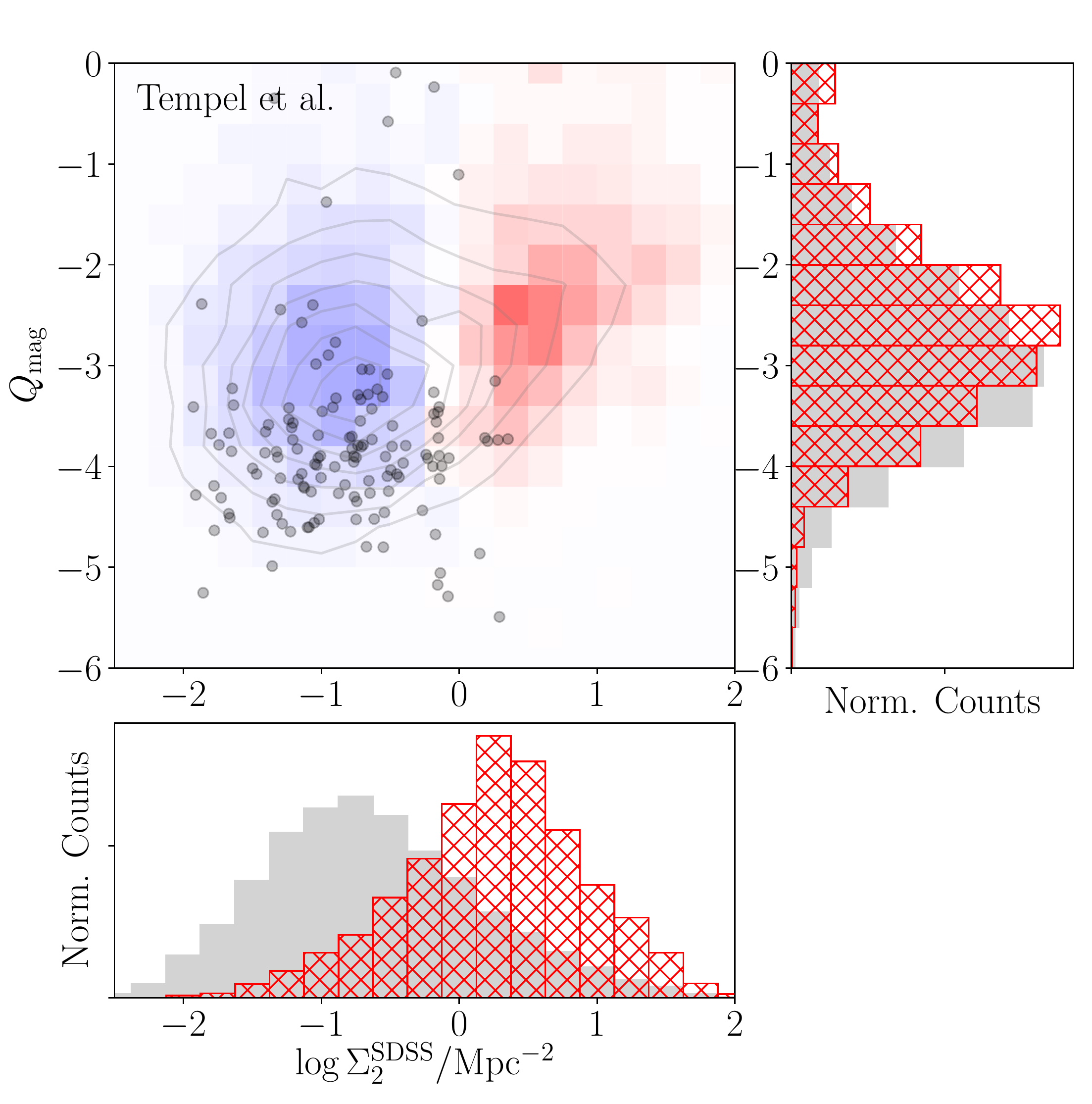}
\includegraphics[width=\columnwidth]{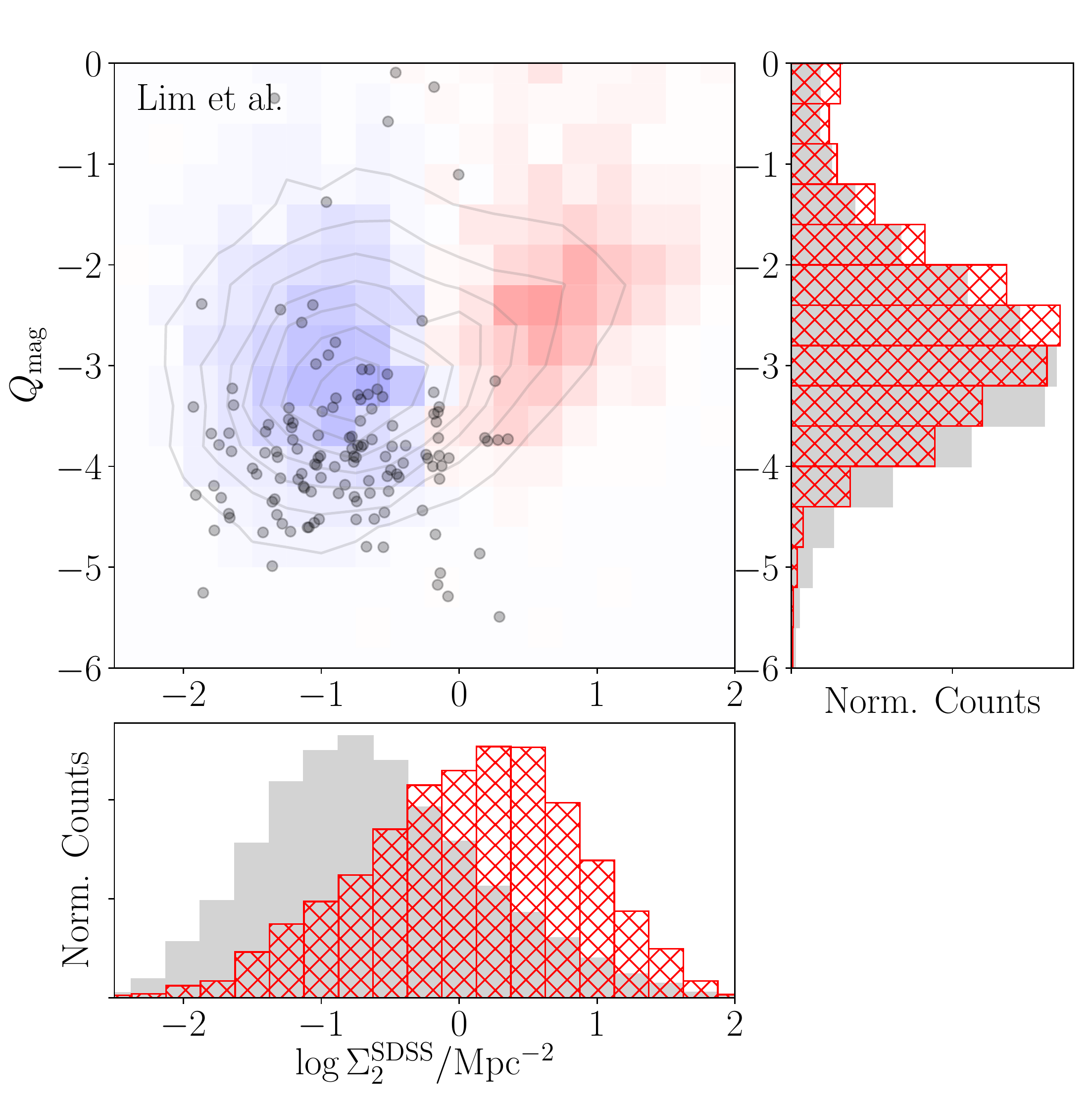}
\caption{The 2nd nearest neighbour density and tidal $Q$ parameter distributions as measured with the SDSS spectroscopic and photometric reference catalogues. The side panels on each of the four plots show the 1D distributions for the group galaxies (red cross-hatched bars) and the whole ALFALFA population (grey filled bars) over the same redshift range as the group catalogue. In the main panels the grey contours indicate the 2D distribution of the ALFALFA population and the grey points are isolated galaxies from the AMIGA sample. The coloured 2D histogram indicates the ratio of the (normalised) histograms of the ALFALFA and group samples. It is blue when the ALFALFA distribution is higher, red when it is lower, and white when equal.}
\label{fig:group_env}
\end{figure*}

To further quantify and compare the environment of the ALFALFA galaxies assigned to the different group catalogues we calculated the projected 2nd nearest neighbour density ($\Sigma_{2}$) of all ALFALFA sources (in the range 1500 $<$ $cz_{\mathrm{cmb}}$/\kms \ $<$ 15000) following the methodology of \citet{Jones+2016b}. For this we used a volume-limited catalogue of SDSS spectroscopic sources \citep{SDSSDR14} within the redshift range 1000 $<$ $cz_{\mathrm{cmb}}$/\kms \ $<$ 15500. Only primary SDSS objects identified as galaxies with clean photometry are included. Each ALFALFA galaxy has an exclusion zone of 5\arcsec \ to prevent self-neighbours. 

This type of environment metric is very common, but we also make use of a photometric metric of environment, the tidal $Q$ parameter \citep{Dahari1984,Verley+2007,Argudo-Fernandez+2014}, because this completely avoids problems of fibre collisions in the SDSS spectroscopic sample and includes galaxies below the brightness cutoff for the spectroscopic survey. The $Q$ parameter is a logarithmic measure of the ratio of the strength of a galaxy's internal gravitational binding forces to the strength of external tidal forces due to its neighbours. 

We follow the approach of \citet{Argudo-Fernandez+2014}, using $r$-band magnitude as a proxy for the total mass of each galaxy, and we use the SDSS $r$-band Petrosian radii as the estimate of their size. This gives the tidal $Q$ parameter as
\begin{equation}
    Q_{\mathrm{mag}} = \log \sum_{i} \frac{D_{p}^{3}}{R_{ip}^{3}} 10^{0.4(m_{r}^{p} - m_{r}^{i})}
\end{equation}
where $D_{p}$ is the diameter of the target object, $R_{ip}$ is the projected separation between the target and the $i$th neighbour, $m_{r}^{p}$ is the $r$-band magnitude of the target, and $m_{r}^{i}$ is the $r$-band magnitude of the $i$th neighbour. We consider all SDSS photometric neighbours within a separation of 80 times the diameter of each target object, that also have radii within a factor of 4 of the radius of the target. The latter criterion is intended to limit the calculation to galaxies at a similar distance as the target.\footnote{The exact choice of values used in calculating this metric are somewhat arbitrary, but are chosen to be as consistent as possible with existing literature.} As before, the reference SDSS catalogue requires the sources to be primaries, identified as galaxies, have clean photometry, and in addition, have radii larger than 2\arcsec \ and photometric redshift estimates of less than 0.1. The additional constraints are designed to eliminate remaining stars and galaxies well beyond the redshift range of our sample. Even with these constraints the disadvantage of a purely photometric definition of enviornment is that interlopers are impossible to completely remove.

The distribution of these environment metrics for the ALFALFA objects in groups (with the same cuts as discussed in \S\ref{sec:group_cats}) compared to the full ALFALFA sample (over the same redshift range as each group catalogue) is shown in Fig. \ref{fig:group_env}. The red/blue shaded histograms indicate the ratio of the normalised distributions, that is, the red shading indicates regions where the normalised distribution is higher for the group galaxies, blue indicates it is higher for general ALFALFA galaxies, and white indicates approximate equality. Here we see that for all four group catalogues the ALFALFA group galaxies appear to be focused in the same region of the parameter space in the upper right quadrant of each panel. The grey points in the Fig. \ref{fig:group_env} indicate the location of the isolated galaxies from the AMIGA \citep{Verdes-Montenegro+2005} sample \citep[a revision of the catalogue of isolated galaxies,][]{Karachentseva1973} that meet the isolation criteria of \citet{Verley+2007}. These are shown for comparison and show minimal overlap with the region where the group galaxies are concentrated, although there are some outliers with high $Q_\mathrm{mag}$ values which are likely the result of interlopers in the calculation of the metric. It should be noted that the environment metrics for the AMIGA galaxies have been re-calculated here as our definitions differ slightly from those of \citet{Verley+2007}.

This comparison is encouraging in that it indicates that based on 2 commonly used environment metrics the four group catalogues are all identifying comparable environments, and so the comparison of the HIMF in each is warranted. However, the reverse is also true, that this indicates caution is needed when using such metrics as they are not sufficiently precise to highlight the subtle differences which we know exist between these group catalogues.

The one dimensional distributions of the projected 2nd nearest neighbour density ($\Sigma_{2}$) indicates that the group galaxies are mostly concentrated in the two upper quartiles, which would lead one to expect a flattening of the low-mass slope in those quartiles of $\Sigma_{2}$, relative to the lowest quartiles, yet this was not seen by \citet{Jones+2016b}. The likely explanation is that the normalised counts in Fig. \ref{fig:group_env} give a somewhat misleading representation of what is happening. While the full ALFALFA comparison sample (grey bars and contours in Fig. \ref{fig:group_env}) typically has around 15000 (in the Spring sky) high signal-to-noise sources within the redshift range covered by the group catalogues, the number of ALFALFA high signal-to-noise sources assigned to the groups is roughly 1500-3000. Thus, there is no region of the parameter space where the group galaxies dominate in number over the field galaxies in ALFALFA. Therefore, all four quartiles of environment in \citet{Jones+2016b} will have been dominated by field galaxies, making the null result regarding the flattening of the low-mass slope unsurprising.

\section{Source confusion in groups}
\label{sec:confusion}

While confusion has been found not to be a concern for studies of the global HIMF \citep{Jones+2015}, the group galaxy HIMF presents quite a different scenario where the typical galaxy number density is much higher. The approach used by \citet{Jones+2015}, based on the 2 point correlation function of HI-selected galaxies, is not appropriate in this case both because we have selected a particular environment and because many of the galaxies in groups may not be as rich in \hi \ as a typical HI-selected galaxy. 

To address the potential issue of confusion we employed a different approached based on the group catalogue itself, specifically the \citetalias{Tempel+2014} catalogue. Using the SDSS DR14 \citep{SDSSDR14} photometry of all optical galaxies in each group, we estimated their velocity widths ($W_{50}$) via the $i$-band Tully-Fisher relation (TFR) of \citep{Ponomareva+2017}. For each ALFALFA \hi \ detection included in the group galaxy HIMF we counted the number of neighbours within 3.5\arcmin \ (approximately the HPBW of Arecibo at 21 cm) that have overlapping \hi \ line emission based on their SDSS redshifts and $W_{50}$ estimates from the TFR (assuming the spectral profiles are top-hat functions). An exclusion zone of 20\arcsec \ and 70 \kms \ is placed around the ALFALFA source in question to avoid self neighbours. This proceedure indicated that 79 per cent of the ALFALFA sources in the \citetalias{Tempel+2014} groups should be free of confused emission, about 16 per cent are potentially confused with another galaxy, and the remaining 5 per cent are potentially confused with multiple other galaxies.

To estimate how severe the confusion is for the $\sim$20 per cent of sources that are potentially confused we used the scaling relation of \citet{Brown+2015} between stellar surface density ($\mu_\ast$) and \hi \ gas fraction, fitting a straight line to their data points (their Table 1) to get the relation: $\log (M_{\mathrm{H\,\textsc{i}}}/M_{\ast}) = 6.842 - 0.872\log(\mu_{\ast}/\mathrm{M_{\odot}\,kpc^{-2}})$. For the stellar mass estimates we used the absolute $i$-band magnitude and the $g-i$ colour as described in \citet{Taylor+2011}. Each of the potentially confused neighbours found in the previous step is thus assigned an expected \hi \ mass. We then estimate the amount of confusion from each neighbour assuming they are point sources and using a Gaussian beam response of FWHM of 3.5\arcmin. The amount of emission is further weighted by the fraction of the full velocity width of each neighbour's profile that is overlapping with the target source's profile, again assuming top-hat profiles. 

Based on these estimates the median amount of fractional excess emission due to confusion is approximately 10 per cent, for the sources which are potentially confused. However, there are a number of outliers with estimates of more than 100 per cent fractional excess due to confusion. Having said this, these outliers make up a total of 1.3 per cent of the ALFALFA sources in the group catalogue. Furthermore, a large fraction of these are only confused with one neighbour, probably indicating that these are in fact self matches and the exclusion zone was insufficient either due to a mismatch in the \hi \ and optical redshifts or the optical centre in SDSS versus that identified manually in ALFALFA.

In summary, analysis of the \citetalias{Tempel+2014} groups indicate that approximately 80 per cent of the ALFALFA sources should not suffer from any confusion, with the remaining 20 per cent typically expected to experience a fractional increase in \hi \ flux of 10 per cent due to confusion. Hence, we do not expect confusion to significantly impact our measurement of the group galaxy HIMF and it is therefore neglected in our method.

\section{Group assignment tables}
\label{sec:group_tabs}

Tables \ref{tab:berlind_groups_tab} to \ref{tab:lim_groups_tab} show the groups that ALFALFA detections are assigned to in this work (complete versions are available in the online version of the article). The columns are as follows:
\begin{itemize}
    \item \textit{Column 1}: AGC identifier of each galaxy as in \citet{Haynes+2018}.
    \item \textit{Column 2}: Group ID number for each group in the respective catalogues (section \ref{sec:group_cats}).
    \item \textit{Column 3}: Group distance assuming pure Hubble-Lema\^{i}tre flow and $H_0 = 70 \; \mathrm{km\,s^{-1}\,Mpc^{-1}}$.
    \item \textit{Column 4}: Logarithm of galaxy \hi \ mass (in solar masses) assuming the group distance.
    \item \textit{Column 5}: Velocity width of the \hi \ spectral line profile at the 50 per cent level in \kms \ \citep{Haynes+2018}.
    \item \textit{Column 5}: Flag set to 1 if the ALFALFA detection falls above the 50 per cent completeness limit for ``Code 1'' sources \citep{Haynes+2011}.
    \item \textit{Column 6}: Flag set to 1 if the ALFALFA detection was matched to the group due to its proximity in phase space instead of as a direct counterpart match to a group member in the original catalogue (section \ref{sec:prox_match}).
\end{itemize}
The full tables also included groups with no ALFALFA detections if the groups fall within the survey volume considered. In these cases the columns corresponding to the \hi \ properties of the galaxies are left blank.

\begin{table*}
\centering
\caption{ALFALFA galaxies assigned to Berlind et al. groups}
\label{tab:berlind_groups_tab}
\begin{tabular}{ccccccc}
\hline\hline
AGC  & Group ID & $D\,h_{70}/\mathrm{Mpc}$ & $\log (M_\mathrm{HI}h_{70}^{2}/\mathrm{M_\odot})$ & $W_{50}/\mathrm{km\,s^{-1}}$ & Completeness Limit & Proximity Match \\ \hline
4061 & 6590     & 116.8                    & 10.11                                             & 489                       & 1                  & 0               \\
4155 & 6985     & 120.3                    & 10.14                                             & 403                       & 1                  & 0               \\
4156 & 6985     & 120.3                    & 10.03                                             & 513                       & 1                  & 0               \\
4211 & 19725    & 151.1                    & 10.18                                             & 476                       & 1                  & 0               \\
4216 & 19063    & 160.3                    & 10.22                                             & 421                       & 1                  & 0               \\ \hline
\end{tabular}
\end{table*}

\begin{table*}
\centering
\caption{ALFALFA galaxies assigned to Yang et al. groups}
\label{tab:yang_groups_tab}
\begin{tabular}{ccccccc}
\hline\hline
AGC  & Group ID & $D\,h_{70}/\mathrm{Mpc}$ & $\log (M_\mathrm{HI}h_{70}^{2}/\mathrm{M_\odot})$ & $W_{50}/\mathrm{km\,s^{-1}}$ & Completeness Limit & Proximity Match \\ \hline
3995 & 5429     & 70.0                    & 9.85                                             & 447                       & 1                  & 0               \\
4109 & 3598     & 197.4                    & 10.46                                             & 304                       & 1                  & 0               \\
4123 & 5484     & 70.6                    & 9.87                                             & 415                       & 1                  & 0               \\
4155 & 2591    & 118.9                    & 10.14                                             & 403                       & 1                  & 0               \\
4156 & 2591    & 118.9                    & 10.03                                             & 513                       & 1                  & 0               \\ \hline
\end{tabular}
\end{table*}

\begin{table*}
\centering
\caption{ALFALFA galaxies assigned to Tempel et al. groups}
\label{tab:tempel_groups_tab}
\begin{tabular}{ccccccc}
\hline\hline
AGC  & Group ID & $D\,h_{70}/\mathrm{Mpc}$ & $\log (M_\mathrm{HI}h_{70}^{2}/\mathrm{M_\odot})$ & $W_{50}/\mathrm{km\,s^{-1}}$ & Completeness Limit & Proximity Match \\ \hline
4061 & 908      & 116.8                    & 10.11                                             & 489                       & 1                  & 0               \\
4123 & 160      & 70.4                    & 9.87                                             & 415                       & 1                  & 0               \\
4145 & 256      & 70.3                    & 9.40                                             & 410                       & 1                  & 0               \\
4154 & 695     &  68.3                    & 9.51                                             & 260                       & 1                  & 0               \\
4155 & 616     & 119.5                    & 10.14                                             & 403                       & 1                  & 0               \\ \hline
\end{tabular}
\end{table*}

\begin{table*}
\centering
\caption{ALFALFA galaxies assigned to Lim et al. groups}
\label{tab:lim_groups_tab}
\begin{tabular}{ccccccc}
\hline\hline
AGC  & Group ID & $D\,h_{70}/\mathrm{Mpc}$ & $\log (M_\mathrm{HI}h_{70}^{2}/\mathrm{M_\odot})$ & $W_{50}/\mathrm{km\,s^{-1}}$ & Completeness Limit & Proximity Match \\ \hline
3969 & 26527    & 119.0                    & 9.90                                             & 374                       & 1                  & 0               \\
4061 & 3329     & 117.0                    & 10.11                                             & 489                       & 1                  & 0               \\
4099 & 8075     & 70.7                    & 9.91                                             & 305                       & 1                  & 0               \\
4123 & 8075    & 70.7                    &  9.87                                             & 415                       & 1                  & 0               \\
4145 & 1290    & 70.2                    & 9.40                                             & 410                       & 1                  & 0               \\ \hline
\end{tabular}
\end{table*}


\bsp	
\label{lastpage}
\end{document}